\documentclass{article}
\usepackage{amsmath,amssymb,amsfonts,amsthm}
\usepackage{graphicx} 
\usepackage[dvipsnames]{xcolor}
\usepackage{tikz}

\newcommand{\tr}{\mathrm{tr}}
\newcommand{\dif}{\mathrm{d}}

\theoremstyle{plain}
\newtheorem{thm}{Theorem}[section]
\newtheorem{lemma}[thm]{Lemma}

\newtheorem{cor}[thm]{Corollary}
\theoremstyle{definition}
\newtheorem{dfn}[thm]{Definition}

\newtheorem{remark}[thm]{Remark}

\numberwithin{equation}{section}

\usepackage{hyperref}

\title{Correlators of the Jacobi Unitary Ensemble on a General Interval and Hurwitz Numbers}
\author{Matvey Sushkov$^{1}$\\
{\small$^{1}$ 
  St. Petersburg Department
    of Steklov Mathematical Institute}\\
  {\small of Russian Academy of Sciences}\\
  {\small 191023, Fontanka 27, St. Petersburg, Russia}\\
{\small email: matveysushkov34@gmail.com}}

\date{June 2026}

\begin{document}

\maketitle

\begin{abstract}
    We study correlators of the Jacobi Unitary Ensemble $\mathbf J(p,q,\alpha,\beta,N)$
    of Hermitian matrices with spectrum in an arbitrary real interval $[p,q]$.
    For $(p,q)=(0,1)$, Gisonni, Grava and Ruzza expressed the large-$N$
    expansion of the correlators $\langle\prod_j \mathrm{tr}\,X^{\lambda_j}\rangle$
    through triple monotone Hurwitz numbers. We extend this genus expansion to
    arbitrary $p,q$ with $\alpha/N$ and $\beta/N$ fixed. Using the Frobenius characteristic
    map and the idempotent basis of the center of $\mathbb C[\mathfrak S_n]$,
    we also obtain an explicit formula for the Schur--Selberg type integral
    $\int_{(p,q)^N}\chi_\lambda(x)\,\Delta^2(x)\prod_a (q-x_a)^\beta (x_a-p)^\alpha\,\mathrm{d}^Nx$
    in terms of irreducible characters of the symmetric group and content products. As consequences, we describe the zero-mean ensemble, and we recover, term by term in the genus expansion, the polygon-gluing
    formula for the Gaussian Unitary Ensemble in the limit $c_A=c_B\to\infty$.
    Structurally, passing from $[0,1]$ to $[p,q]$ leaves the hypergeometric weight of the
    associated 2D Toda tau function unchanged and only changes the specialization of its second set of times.
\end{abstract}

\section{Introduction and main results}
\label{sec:intro}

\subsection{The Jacobi Unitary Ensemble and correlators on a general interval}

We denote by $\mathcal{H}_N([p, q])$ the set of $N \times N$ Hermitian matrices whose eigenvalues lie in the interval $[p, q]$. We endow this space with the Lebesgue measure
\begin{equation}
    \dif X = \prod_{i=1}^N \dif X_{ii} \prod_{1 \le j < k \le N} \dif \mathrm{Re}X_{jk} \,\dif \mathrm{Im} X_{jk}.
\end{equation}
The Jacobi Unitary Ensemble on $[p, q]$ is then defined as the probability measure on $\mathcal{H}_N([p, q])$ given by
\begin{equation}
    \dif \mu (X) = \frac{1}{C} \det{}^\alpha (X - p I) \det{}^\beta (qI - X) \dif X,
\end{equation}
where the parameters $\alpha$ and $\beta$ obey the conditions $\mathrm{Re}\,\alpha, \mathrm{Re}\,\beta > -1$. The normalization constant $C$ is given by 
\begin{equation}
    \begin{aligned}
        & C = \int_{\mathcal{H}_N([p,q])} \det{}^\alpha (X - pI) \det{}^\beta (q I - X) \dif X = \\
        & = \pi^{N(N-1)/2}\,(q-p)^{N(\alpha+\beta+N)}\prod_{j=1}^{N}\frac{\Gamma(\alpha+j)\,\Gamma(\beta+j)}{\Gamma(\alpha+\beta+N+j)} .
    \end{aligned}
\end{equation}
It ensures that $\dif \mu$ has total mass 1. We will denote this ensemble as $\mathbf{J}(p,q,\alpha,\beta, N)$. 

\begin{remark}[Matrix model for integer parameters]
When $\alpha,\beta\in\mathbb Z_{\ge0}$, set $M_A=N+\alpha$, $M_B=N+\beta$, let $H_A,H_B$ be complex Gaussian
random matrices of size $M_A\times N$ and $M_B\times N$, and put $A=H_A^\dagger H_A$, $B=H_B^\dagger H_B$ (Wishart matrices).
Then $(A+B)^{-1/2}(qA+pB)(A+B)^{-1/2}$ is distributed according to $\mathbf J(p,q,\alpha,\beta,N)$ for $p<q$.
This realization exists only for non-negative integer $\alpha,\beta$. It is not used in the proofs and serves as motivation.
\end{remark}

\paragraph{Notation.}
Throughout, $N$, $\alpha$ and $\beta$ are fixed, and $c_A=1+\alpha/N$, $c_B=1+\beta/N$.
We write $X_{0,1}$ for a random matrix with law $\mathbf J(0,1,\alpha,\beta,N)$, and for arbitrary real $p,q$ we define
\begin{equation}
    \label{eq:Xpq-def}
    X_{p,q}:=(q-p)\,X_{0,1}+pI.
\end{equation}
The symbol $\langle\,\cdot\,\rangle$ denotes the expectation with respect to the law of the matrix appearing inside the brackets.
By Lemma~\ref{lem:affine} below, this definition is consistent with the density above: for $p<q$ the matrix $X_{p,q}$ has law $\mathbf J(p,q,\alpha,\beta,N)$, for $p>q$ it has law $\mathbf J(q,p,\beta,\alpha,N)$, and for $p=q$ it is the deterministic matrix $pI$.

The probability density function of eigenvalues of matrices in $\mathbf{J}(p,q,\alpha,\beta, N)$ is 
\begin{equation}
    \frac{1}{\tilde{C}}\prod_{i=1}^N (q-y_i)^\beta (y_i - p)^\alpha \prod_{j<k} (y_k - y_j)^2,
\end{equation}
where 
\begin{equation}
    \tilde C =(q-p)^{N(\alpha+\beta+N)}\prod_{j=1}^{N}\frac{\Gamma(\alpha+j)\Gamma(\beta+j)\Gamma(j+1)}{\Gamma(\alpha+\beta+N+j)}.
\end{equation}

Our main focus is on the following expectation values:
\begin{equation}
    \label{eq:bracket}
    \left<\prod_{j=1}^{\ell(\lambda)} \tr X^{\lambda_{j}}\right> 
    = \frac{1}{C}\int_{\mathcal{H}_N([p,q])} \prod_{j=1}^{\ell(\lambda)} \tr X^{\lambda_{j}} \,\det{}^\alpha (X - pI)\, \det{}^\beta (q I - X) \dif X,
\end{equation}
where the $\lambda_j$ are positive integers satisfying $\lambda_1 \ge \lambda_2 \ge \cdots$. We refer to the $\lambda_j$ as the parts of the partition $\lambda = (\lambda_1, \ldots, \lambda_{\ell(\lambda)})$, whose length (i.e. number of parts) is denoted by $\ell(\lambda)$.

Let $\mathfrak{S}_n$ be the symmetric group on $n$ symbols. Given a partition $\lambda$ that specifies a cycle type, we denote by $[\lambda]$ the conjugacy class consisting of all permutations with cycle type $\lambda$. We write $w_\lambda$ for a chosen representative of this conjugacy class. The order of the centralizer of an element of type $\lambda$ is given by
\begin{equation}
    \label{centralizer}
    z_\lambda = \prod_{i\ge1} i^{m_i} m_i !,
\end{equation}
where $m_i$ counts how many parts of $\lambda$ are equal to $i$. Consequently, the size of the conjugacy class $[\lambda]$ is
\begin{equation}
    |[\lambda]| = \frac{n!}{z_\lambda}.
\end{equation}

In their paper \cite{gisonni2021jacobi}, Massimo Gisonni, Tamara Grava, and Giulio Ruzza used Young–Jucys–Murphy elements together with the Schur–Selberg integral to obtain the following result. For $p = 0, q = 1$ they proved that
\begin{equation}
    \label{GGR-correlators}
    \begin{aligned}
        & \left<\prod_{j=1}^{\ell(\lambda)}\tr X^{\lambda_{j}}_{0,1}\right> = \\
        & = \sum_{g\in \mathbb{Z}}\frac{(-1)^{|\lambda|}}{N^{2g-2+\ell(\lambda)}}\sum_{\mu,\nu\vdash|\lambda|}\frac{c_{A}^{\ell(\nu)}\, h_{g}(\lambda,\mu,\nu)}{(-c_{A}-c_{B})^{\ell(\mu)+\ell(\nu)+\ell(\lambda)+2g-2}\, |[\lambda]|},
    \end{aligned}
\end{equation}
where the number $h_g(\lambda, \mu, \nu)$ is defined as follows.

\begin{dfn}
    \label{def:h_g}
    Given $n \geq 0$, three partitions $\lambda, \mu, \nu \vdash n$ and an integer $g$, we define $h_g(\lambda, \mu, \nu)$ to be the number of tuples $\left(w_\mu, w_\nu, \tau_1, \ldots, \tau_r\right)$ of permutations in $\mathfrak{S}_n$ such that
    \begin{enumerate}
        \item $r=2 g-2-n+\ell(\mu)+\ell(\nu)+\ell(\lambda)$,
        \item $w_\mu \in [\mu], w_\nu \in [\nu]$,
        \item $\tau_i=\left(a_i, b_i\right)$ are transpositions, with $a_i<b_i$ and $b_1 \leq \cdots \leq b_r$, and
        \item $w_\mu w_\nu \tau_1 \cdots \tau_r \in [\lambda]$.
    \end{enumerate}
\end{dfn}

\begin{remark}[Geometric meaning of $h_g$ and of the genus]
\label{rem:hurwitz-geometry}
Put $\sigma = w_\mu w_\nu \tau_1\cdots\tau_r \in [\lambda]$. The tuple
$(w_\mu, w_\nu, \tau_1,\dots,\tau_r, \sigma^{-1})$ is a tuple of permutations in $\mathfrak S_n$
with product equal to the identity. By the Riemann existence theorem, such a tuple
(up to simultaneous conjugation) is the monodromy of a degree-$n$ branched covering of $\mathbb{CP}^1$.
The covering has ramification profiles $\mu,\nu,\lambda$ over three marked points
(say $0,1,\infty$) and $r$ further simple branch points. The covering surface
need not be connected. For a possibly disconnected surface $\Sigma$ we define $g$ by
$\chi(\Sigma)=2-2g$, so that $2-2g=\sum_i(2-2g_i)$ over its components.
Hence $g$ may be zero or negative, which is why the sum in Theorem~\ref{thm:main-theorem}
runs over $g\in\mathbb Z$. The Riemann--Hurwitz formula
\[
2g-2=-2n+\bigl(n-\ell(\mu)\bigr)+\bigl(n-\ell(\nu)\bigr)+\bigl(n-\ell(\lambda)\bigr)+r
\]
is exactly the relation $r=2g-2-n+\ell(\mu)+\ell(\nu)+\ell(\lambda)$ in item~1 of the definition.
The condition $b_1\le\dots\le b_r$ on the transpositions is a monotonicity condition.
Thus $h_g$ is a triple \emph{monotone} Hurwitz number in the sense of Goulden--Guay-Paquet--Novak \cite{goulden2012polynomialitymonotonehurwitznumbers, Goulden_2013}. It counts monotone factorizations, not coverings weighted by $1/|\mathrm{Aut}|$. However, the genus attached to every factorization is the genus of the
corresponding covering. For this reason we may speak of the genus expansion.
\end{remark}

The aim of this paper is to generalize the result \eqref{GGR-correlators} to matrices in $\mathbf{J}(p,q,\alpha,\beta, N)$ for arbitrary $p$ and $q$. We prove the following theorem.

\begin{thm}
    \label{thm:main-theorem}
    Let $p,q\in\mathbb R$ and let $X_{p,q}$ be the random matrix defined in \eqref{eq:Xpq-def}. Let $\lambda$ be an arbitrary partition of a positive integer $n$. Then, for fixed $c_A$ and $c_B$ (that is, $\alpha=(c_A-1)N$, $\beta=(c_B-1)N$), the correlators of these matrices admit the following Laurent expansion in $N^{-1}$ as $N\to\infty$:
    \begin{equation}
        \label{eq:theorem_statement}
        \begin{aligned}
            & \left<\prod_{j=1}^{\ell(\lambda)}\tr X_{p,q}^{\lambda_{j}}\right> = \sum\limits_{g \in \mathbb{Z}} \frac{(-1)^{|\lambda|}}{N^{2g-2+ \ell(\lambda)}}\cdot\\
            & \cdot \sum\limits_{\mu, \nu \vdash |\lambda|} \frac{\prod_{i=1}^{\ell(\nu)}\left( c_A q^{\nu_{i}} + c_B p^{\nu_{i}} \right)}{(-c_{A} - c_{B})^{2g-2 + \ell(\lambda) + \ell(\mu) + \ell(\nu)}} \frac{h_{g}(\lambda, \mu, \nu)}{|[\lambda]|}, 
        \end{aligned}
    \end{equation}
    where the number $h_g(\lambda, \mu, \nu)$ is the number defined in Definition \ref{def:h_g}.
\end{thm}

It should be emphasized that the parameters $(p,q)$ are best interpreted as indicating a direction in parameter space, rather than fixing an absolute normalization. In fact, by \eqref{eq:Xpq-def}, $X_{kp,kq}=kX_{p,q}$ for every real $k$, hence rescaling to $(kp,kq)$ multiplies the correlator by $k^{|\lambda|}$. This is visible in \eqref{eq:theorem_statement}: the parameters $p,q$ enter only through the factor $\prod_{i}(c_Aq^{\nu_i}+c_Bp^{\nu_i})$, which is multiplied by $k^{\sum_i\nu_i}=k^{|\lambda|}$.

Several special choices of $(p,q)$ are discussed in
Section~\ref{sec:particular-p-q}: the degenerate case $p=q$, the standard
Jacobi ensemble, the zero-mean ensemble, and the Gaussian limit.

\subsection{Partition function and Schur–Selberg integral on $[p,q]$}

We define the partition function of the JUE correlators by
\begin{equation}
    \label{eq:partition_function}
    Z_N(\mathbf{u})
    :=
    \int_{\mathcal{H}_N([p,q])}
    \exp\left(
    \sum_{k\geq 1}\frac{u_k}{k}\operatorname{tr}X^k
    \right)
    \,\dif\mu(X).
\end{equation}
Equivalently, expanding the exponential in powers of the parameters $\mathbf{u}=(u_1,u_2,\ldots)$, we obtain the formal power series
\begin{equation}
    \label{eq:pertfunc power sums}
    Z_N(\mathbf{u})
    =
    \sum_{\lambda\in\mathcal{P}}
    \frac{
    \left\langle
    \prod_{j=1}^{\ell(\lambda)}
    \tr X^{\lambda_j}
    \right\rangle
    }{z_\lambda}
    \,p_\lambda(\mathbf{u}),
\end{equation}
where $\mathcal{P}$ denotes the set of all partitions, $z_\lambda$ is given in \eqref{centralizer} and
\begin{equation}
p_\lambda(\mathbf{u})
:=
\prod_{i=1}^{\ell(\lambda)}u_{\lambda_i}
\end{equation}
is a power-sum symmetric function. 

The same partition function can be written in the basis of Schur symmetric functions $s_\lambda(\mathbf{u})$.
\begin{equation}
    \label{eq:partfunc schur func}
    Z_N(\mathbf{u}) = \sum_{\lambda, \ell(\lambda) \le N} c_{\lambda, N}  s_{\lambda}(\mathbf{u}).
\end{equation}

Let us introduce the integral
\begin{equation}
\label{eq:Schur-Selberg-like-integral}
    I_{\lambda} = \int_{(p,q)^{N}} \chi_{\lambda}(x) \Delta^{2}(x)\prod_{a=1}^{N}(q-x_{a})^{\beta} (x_{a}-p)^{\alpha} \dif^{N}x,
\end{equation}
where $\Delta(x)$ is the Vandermonde determinant and the functions $\chi_\lambda(x)$ are the characters of $\mathrm{GL}_N$ given by 
\begin{equation}
    \chi_\lambda (x) = \frac{\det \left[x_j^{N-i+\lambda_i}\right]_{i,j=1}^{N}}{\Delta(x)}.
\end{equation}
These characters are also Schur polynomials (not to be confused with Schur functions of infinitely many arguments $s_\lambda(\mathbf{u})$).

By Lemma~2.3 of \cite{gisonni2021jacobi}, which is valid for an arbitrary potential on an arbitrary interval, the coefficients $c_{\lambda,N}$ of the expansion \eqref{eq:partfunc schur func} satisfy
\begin{equation}
    \label{eq:clambda by Ilambda}
    c_{\lambda,N} = \frac{I_\lambda}{I_\varnothing},
\end{equation}
where $\varnothing$ denotes the empty partition, $c_{\lambda,N}=0$ for $\ell(\lambda)>N$, and the weight is $(q-x)^\beta(x-p)^\alpha$ on $(p,q)$. The proof of that lemma is the Cauchy identity
\[
\exp\Bigl[\sum_{k\ge1}\frac{u_k}{k}\,(x_1^k+\dots+x_N^k)\Bigr]=\sum_{\ell(\lambda)\le N}\chi_\lambda(x)\,s_\lambda(\mathbf u),
\]
integrated over the eigenvalues $x_1,\dots,x_N$.

The integral \eqref{eq:Schur-Selberg-like-integral} for $(p,q) = (0,1)$ is the well-known Schur-Selberg integral \cite[page 385]{macdonald1995symmetric}. The last integral can be calculated as follows.
\begin{equation}
    \label{eq:01-schur-selberg}
    \begin{aligned}
        & \int_{(0,1)^{N}} \chi_{\lambda}(x) \Delta^{2}(x)\prod_{a=1}^{N}(1-x_{a})^{\beta} x_{a}^{\alpha} \dif^{N}x = \\
        & = N!\prod_{1 \leq i<j \leq N}\left(\lambda_i-\lambda_j+j-i\right) \prod_{k=1}^N \frac{\Gamma(\beta+k) \Gamma\left(\alpha+N+\lambda_k-k+1\right)}{\Gamma\left(\alpha+\beta+2 N+\lambda_k-k+1\right)},
    \end{aligned}
\end{equation}

The value of the integral $I_\lambda$ for arbitrary $(p,q)$ is given by the following theorem, which will be proved in this paper.

\begin{thm}
    \label{thm:schur-selberg}
    With the notation given above, for $\varepsilon = N^{-1}$, for given parameters $c_A,c_B$, endpoints $p < q$ and for given partition $\lambda \vdash n$ the following holds.
    \begin{equation}
        \label{eq:theorem_schur-selberg-type}
        \begin{aligned}
            &\frac{I_\lambda}{I_{\varnothing}}  = \left(\prod_{(i,j) \in \lambda} \frac{1+ \varepsilon (j-i)}{1+ \frac{\varepsilon}{c_{A}+c_{B}} (j-i)}\right) \cdot \\
            & \cdot \sum\limits_{\nu \vdash n} \frac{\chi^{\lambda}(\nu)}{z_{\nu}} \frac{\varepsilon^{-\ell(\nu)}}{(c_{A}+c_{B})^n}\prod_{k=1}^{\ell(\nu)} (q^{\nu_{k}} c_{A}+ p^{\nu_{k}}c_B),
        \end{aligned}
    \end{equation}
    where $\chi^{\lambda}(\cdot)$ is the irreducible character corresponding to the representation $V^\lambda$ of the symmetric group $\mathfrak{S}_n$.
    The normalization constant $I_\varnothing$ is given by
    \begin{equation}
        I_{\varnothing} = (q-p)^{N(\alpha+\beta+N)} \prod_{j=0}^{N-1} \frac{\Gamma(\alpha+j+1)\, \Gamma(\beta+j+1)\, \Gamma(j+2)}{\Gamma(\alpha+\beta+N+j+1)}.
    \end{equation}
\end{thm}

Clearly, for general $p,q$ the integral $I_\lambda$ can be evaluated by the
substitution $x_a=(q-p)y_a+p$. It maps $(p,q)^N$ to $(0,1)^N$ and gives
$q-x_a=(q-p)(1-y_a)$, $x_a-p=(q-p)y_a$,
$\Delta^2(x)=(q-p)^{N(N-1)}\Delta^2(y)$ and $\dif^N x=(q-p)^N\dif^N y$.
The total power of $(q-p)$ is $N(\alpha+\beta+N)$, hence
\begin{equation}
    I_\lambda=(q-p)^{N(\alpha+\beta+N)}\int_{(0,1)^N}\chi_\lambda\big((q-p)y+p\big)\,
\Delta^2(y)\prod_{a=1}^N(1-y_a)^\beta y_a^\alpha\,\dif^N y .
\end{equation}
For $\lambda=\varnothing$ we have $\chi_\varnothing=1$, and this immediately
gives the normalization constant $I_\varnothing$ from the known $(0,1)$ result.
For general $\lambda$ one uses the "binomial" expansion~\cite[I, §3, Example 10]{macdonald1995symmetric}
\begin{equation}
    \chi_\lambda(ay+b)=\sum_{\mu\subseteq\lambda}a^{|\mu|}b^{|\lambda|-|\mu|}   \det\!\left[\binom{\lambda_i+N-i}{\mu_j+N-j}\right]_{i,j=1}^{\ell(\lambda)}\chi_\mu(y),
\end{equation}
and integrates each $\chi_\mu(y)$ by the $(0,1)$ formula \eqref{eq:01-schur-selberg}. The result is a sum over
all subdiagrams $\mu\subseteq\lambda$, each term containing an $\ell(\lambda)\times \ell(\lambda)$ determinant and a product of $N$ Gamma-function ratios. The dependence on $N$
is hidden in these terms, and extracting the large-$N$ behaviour requires
tracking cancellations between them. In contrast,
\eqref{eq:theorem_schur-selberg-type} depends on $p,q$ only through the
factors $q^{\nu_k}c_A+p^{\nu_k}c_B$ and on $N$ only through the explicit
parameter $\varepsilon$, and it is written in terms of the irreducible
characters $\chi^\lambda(\nu)$, which is what is needed in the combinatorial
setting. Therefore, although $I_\lambda$ can be computed by the change of
variables, we find \eqref{eq:theorem_schur-selberg-type} more practical.
To the best of our knowledge, this formula is new.

To verify formula \eqref{eq:theorem_schur-selberg-type}, the reader can utilize the verification scripts provided in the public repository \url{https://github.com/Abarvg1/JUEGM_source_code.git}. These scripts (developed with the assistance of ChatGPT) compute and compare the ratio $I_\lambda/I_\varnothing$ against the right-hand side of \eqref{eq:theorem_schur-selberg-type}.

\subsection{Motivation and the plan of the paper}

This paper was motivated by our joint work with A.~Nazarov on the limit shapes of Young
diagrams. In that study, we considered the average
diagram with respect to the measure arising from Skew Howe duality \cite{betea2025limit}. The resulting limit shape is the
image, via the Markov--Krein correspondence \cite{kerov1993transition}, of the limit
eigenvalue density of Jacobi Unitary Ensemble matrices.

Our aim was to describe the asymptotics of the largest row of such a diagram
and to relate its statistical properties combinatorially to the largest
eigenvalue of JUE matrices. This parallels the result of Okounkov \cite{okounkov2000random} for the Vershik--Kerov limit shape and the Gaussian Unitary
Ensemble. These expectations were formulated as conjectures in
\cite{nazarov2025randomyoungdiagramsjacobi}. For this purpose it is more convenient to work with
JUE matrices with zero mean, but we could not find results on the
correlators of such matrices in the literature. The paper
\cite{gisonni2021jacobi} derived the correlators for $(p,q)=(0,1)$ from the
Schur--Selberg integral. Taking it as a starting point, we looked for a
generalization to arbitrary $(p,q)$, and that investigation led to the results
presented here.

Concretely, we go in the opposite direction to \cite{gisonni2021jacobi}. We
take their formula for the correlators as input and extend it to arbitrary
$(p,q)$, which gives Theorem~\ref{thm:main-theorem}. We then use the
resulting expression for the partition function to evaluate the
Schur--Selberg type integral on $[p,q]$, which gives
Theorem~\ref{thm:schur-selberg}.

\paragraph{Plan of the paper.}
Section~\ref{sec:prelim} collects the necessary facts about
Young--Jucys--Murphy elements, Hurwitz numbers, central idempotents and the
Frobenius characteristic map. Theorem~\ref{thm:main-theorem} is proved in
Section~\ref{sec:main-thm-proof}. In Section~\ref{sec:particular-p-q} we discuss special choices of
$(p,q)$, including the zero-mean ensemble and its Gaussian limit; these are
placed after the proof because their derivation uses its intermediate
formulas. Theorem~\ref{thm:schur-selberg} is proved in Section~\ref{sec:thm-SS-proof}, and Section~\ref{sec:discussion} contains a discussion and open problems. Sections~\ref{sec:particular-p-q} and~\ref{sec:thm-SS-proof} are independent of each other.

\section{Preliminaries}
\label{sec:prelim}

In this section we assemble the tools needed for the proofs. In Section~\ref{subs:yjm} we review
the class sums and the Young--Jucys--Murphy elements of the group algebra of $\mathfrak S_n$,
show how products of these elements generate the Hurwitz numbers $h_g$, and introduce the
function $G$ that encodes the Jacobi weight. In Section~\ref{subs:idemp} we pass to the
idempotent basis of the center and to the Frobenius characteristic map, which are required
for Theorem~\ref{thm:schur-selberg}.

\subsection{Young–Jucys–Murphy elements and Hurwitz numbers}
\label{subs:yjm}

The group algebra $\mathbb{C}[\mathfrak{S}_n]$ consists of formal linear combinations of
permutations $\sigma\in\mathfrak{S}_n$, with multiplication given by composition of
permutations (we compose from right to left) extended bilinearly. By Maschke's theorem
\cite{Maschke1899} it is semisimple, and by the Artin--Wedderburn theorem
\cite{Wedderburn1908,Artin1927} it is isomorphic to a direct sum of full matrix algebras
over $\mathbb{C}$, one for each irreducible representation $V^\lambda$ of $\mathfrak{S}_n$:
$\mathbb{C}[\mathfrak{S}_n]\cong\bigoplus_{\lambda\vdash n}\mathrm{End}(V^\lambda)$.
The irreducible representations are parametrized by partitions $\lambda\vdash n$
\cite{Frobenius1900,Specht1935}; see also \cite[Ch.~6]{Serre} for the general theory.

The center $Z[\mathbb C[\mathfrak S_n]]$ has two natural descriptions that look completely different.

\begin{itemize}
    \item \emph{Class sums.} Let $C_\mu$ be the sum of all permutations of cycle type $\mu$. The elements $C_\mu$, $\mu\vdash n$, form a basis of the center, so $\dim Z[\mathbb C[\mathfrak S_n]]$ equals the number of partitions of $n$. This basis is natural for combinatorics, since coefficients in this basis count permutations of a given cycle type.
    \item \emph{Young--Jucys--Murphy elements.} Define $\mathcal J_1=0$ and $\mathcal J_k=\sum_{i<k}(i\,k)$, where $(a\,b)$ is the transposition swapping $a$ and $b$. These are commonly referred to as the Jucys--Murphy elements; however, as observed by Vershik and Okounkov in \cite{vershik2005newapproachrepresentationthoery}, they were introduced earlier by Alfred Young. Subsequently, Jucys \cite{jucys1974symmetric} and Murphy \cite{Murphy1983} independently developed the theory of these elements. For this reason we refer to the $\mathcal J_i$ as the Young--Jucys--Murphy elements, or YJM for short. The elements commute pairwise, although they are not themselves central; moreover, they generate the Gelfand--Tsetlin subalgebra \cite{vershik2005newapproachrepresentationthoery}. Jucys \cite{jucys1974symmetric} proved that any symmetric polynomial in $\mathcal J_1,\dots,\mathcal J_n$ lies in the center, and that such polynomials span the entire center. Murphy \cite{murphy1981new} showed that the $\mathcal J_i$ act diagonally in the Young basis with eigenvalues given by the contents of boxes; consequently a symmetric polynomial in the $\mathcal J_i$ acts on $\mathcal E_\lambda$ as the scalar obtained by evaluating it at the contents of $\lambda$. 
\end{itemize}

The two descriptions are related by the following identity, which we use repeatedly.
Fix a constant $\varepsilon$. Then
\begin{equation}
    \label{eq:symmetric-product}
    \prod_{i=1}^n (1+ \varepsilon \mathcal{J}_i) = \sum_{\nu \vdash n} \varepsilon^{n-\ell(\nu)} C_\nu.
\end{equation}
Equivalently, the elementary symmetric polynomial $e_k(\mathcal J_1,\dots,\mathcal J_n)$ is the sum of $C_\nu$
over all $\nu\vdash n$ with $n-\ell(\nu)=k$. Hence every symmetric polynomial in the $\mathcal J_i$ is a linear combination of the class sums $C_\mu$, but for a general symmetric polynomial the coefficients of this expansion are not given by a simple formula. Identity \eqref{eq:symmetric-product} is a notable explicit example: it expands the elementary symmetric polynomials. This identity was originally presented in \cite{jucys1974symmetric}, identity~(8).

Identity~\eqref{eq:symmetric-product} follows from a simple combinatorial fact: every permutation in $\mathfrak{S}_n$ can be written uniquely as a product of transpositions $(a\,b)$ with $a<b$, arranged so that the transpositions are applied in a strictly increasing order (a so-called canonical, or strictly increasing, factorization). Moreover, the number of transpositions required in this factorization is exactly $n-\ell(\nu)$, where $\nu$ is the cycle type of the permutation. Since each factor $\mathcal{J}_i$ contributes one power of $\varepsilon$ precisely when the corresponding transposition is used, expanding the left-hand side of~\eqref{eq:symmetric-product} and collecting terms by the number of transpositions used reproduces the right-hand side.

To make this correspondence transparent, we verify it by constructing an $m$-cycle from transpositions, adding one element at a time.

\begin{enumerate}
    \item \textbf{Inserting a new element into a cycle.}
    Suppose we already have an $(m,1)$-permutation, meaning a permutation consisting of an $m$-cycle together with one fixed point. By multiplying by a suitable transposition, we can merge this fixed point into the cycle, thereby obtaining an $(m+1)$-cycle. The rule is straightforward: to insert a new element $b$ directly after a given element $a$ in the cycle, multiply by the transposition $(a\,b)$. This makes clear how each additional transposition extends the cycle by one element.

    \item \textbf{Constructing an $m$-cycle from transpositions.}
    Repeating the insertion procedure, we can build any $m$-cycle, beginning with a single transposition of two elements, then inserting a third, then a fourth, and so on. Since we start with $2$ elements already combined and add one new element at each step, we need exactly $m-1$ transpositions to obtain a complete $m$-cycle.

    \item \textbf{Recovering the cycle from the product formula.}
    Conversely, let $C = (a_1\, a_2\, \cdots\, a_m)$ be the resulting $m$-cycle, and let $a$ be its smallest element. Consider the product $\prod_i (1+\mathcal{J}_i)$, restricted to the indices $i$ that lie in this cycle. When we expand this product, the term associated with the cycle $C$ is obtained by choosing $1$ from the factor $(1+\mathcal{J}_a)$, and selecting the appropriate transposition from $\mathcal{J}_i$ in each of the remaining factors. Equivalently, this term is precisely the product of the $m-1$ transpositions constructed according to the steps above, which gives the canonical factorization of the $m$-cycle.
\end{enumerate}

Thus, every product of transpositions on the left-hand side of \eqref{eq:symmetric-product} corresponds to a unique permutation on the right-hand side. At the same time, each permutation appears with the prefactor $\varepsilon^{n-\ell(\nu)}$. Permutations sharing the same cycle type can then be collected together into $C_\nu$.

We consider the function
\begin{equation}
    \label{eq:G_func}
    G(t) = \frac{(1+t)(1+\gamma t)}{(1-\delta t)},
\end{equation}
where $\gamma$ and $\delta$ are real parameters. When this function is evaluated on the Young--Jucys--Murphy elements, it recovers the quantities $h_g(\lambda,\mu,\nu)$. We explain this below.

Any element $x$ of the algebra $\mathbb{C}[\mathfrak{S}_n]$ admits an expansion of the form
\begin{equation}
    \label{eq:sym_alg_expansion}
    x = \sum_{\sigma \in \mathfrak{S}_n} x_\sigma \sigma,
\end{equation}
where the coefficients satisfy $x_\sigma \in \mathbb{C}$. Given a specific permutation $\sigma$ and an element $x$ of the algebra, we write $[\sigma]x$ for the corresponding coefficient $x_\sigma$ in the above expansion \eqref{eq:sym_alg_expansion}. In the future we will consider $x_\sigma$ being Laurent series in $N$. In that case $[\sigma N^k]$ means the coefficient of $N^k$ in $x_\sigma$ expansion. If $x$ belongs to the center, then the coefficients $[w_\lambda]x$ and $[\tilde w_\lambda]x$ are the same. It means that we can write $[C_\lambda]x$ instead of $[w_\lambda]x$ in case $x \in Z[\mathbb{C}[\mathfrak{S}_n]]$.

We evaluate the following object
\begin{equation}
    \label{eq:G01_initial_expansion_of_permutations}
    \prod_{i=1}^n G\left(\frac{\mathcal{J}_i}{N}\right) = \prod_{i=1}^n \frac{\left(1+\frac{\mathcal{J}_i}{N}\right) \left(1+ \gamma\frac{\mathcal{J}_i}{N}\right)}{\left(1- \delta\frac{\mathcal{J}_i}{N}\right)},
\end{equation}
where $\left(1-\delta\frac{\mathcal{J}_i}{N}\right)^{-1}$ is understood as formal power series in $N^{-1}$. Using \eqref{eq:symmetric-product}, we obtain
\begin{equation}
    \label{eq:G01_initial_expansion_classes}
    \begin{aligned}
        & \prod_{i=1}^n G\left(\frac{\mathcal{J}_i}{N}\right) =  \left(\sum_{\mu \vdash n}N^{\ell(\mu)-n} C_\mu \right) \left(\sum_{\nu \vdash n} N^{\ell(\nu)-n}\gamma^{n-\ell(\nu)}C_\nu \right) \cdot \\
        & \cdot \left(\sum_{r\ge 0}N^{-r}\delta^r \sum_{1 \le a_1 \le a_2 \le \dots \le a_r \le n}\mathcal{J}_{a_1}\dots \mathcal{J}_{a_r}\right),
    \end{aligned}
\end{equation}

If we expand the parentheses in \eqref{eq:G01_initial_expansion_classes}, then we get a sum of products of the form $w_\mu w_\nu \tau_1 \dots \tau_r$ with factor $N^{-r-2n+\ell(\mu)+\ell(\nu)}$. We consider particular product and denote the result of multiplication as $\sigma$. The genus $g$ of the corresponding (possibly disconnected) covering is determined by the
Riemann--Hurwitz formula, see Remark~\ref{rem:hurwitz-geometry}:
\begin{equation}
    \label{eq:RH_GGR_case}
    \begin{aligned}
        & 2g-2 = -2n+(n-\ell(\nu))+(n-\ell(\mu))+(n-\ell(\sigma))+r = \\
        & = n+r-\ell(\mu)-\ell(\nu)-\ell(\sigma).
    \end{aligned}
\end{equation}
Thus, the total power of $N$ of the product is $-n - \ell(\sigma)-(2g-2)$.

What is the number of products in the expansion of the right-hand side of \eqref{eq:G01_initial_expansion_classes} with fixed $\mu,\nu$ and genus $g$, whose cycle structure is $\lambda$? It is exactly $h_g(\lambda, \mu, \nu)$. As we consider central element, every permutation $w_\lambda$ has the same coefficient. It follows that the number of products equal to particular $w_\lambda$ is $h_g(\lambda, \mu, \nu)/|[\lambda]|$.

Substituting $\gamma=\frac{1}{c_A}$ and $\delta=-\frac{1}{c_A+c_B}$, we obtain
\begin{equation}
    \label{eq:01cor_G_function}
    \left<\prod_{j=1}^{\ell(\lambda)}\tr X^{\lambda_{j}}_{0,1}\right> = \sum\limits_{d\ge 0} N^{|\lambda|-d} \left(\frac{c_{A}}{c_{A}+c_{B}}\right)^{|\lambda|} [w_{\lambda} N^{-d}] \prod_{i=1}^{|\lambda|} G\left(\frac{\mathcal{J}_{i}}{N} \right).
\end{equation}
We note that the factors $c_{A}^{\ell(\nu)}$ and $(-c_{A}-c_{B})^{\ell(\mu)+\ell(\nu)+\ell(\lambda)+2g-2}$ are now the part of function $G$.

\subsection{Central idempotents and the Frobenius characteristic map}
\label{subs:idemp}

So far we have worked in the basis of class sums $C_\nu$, which is the natural
choice for computing correlators. For the Schur--Selberg integral another
basis of the center is more convenient.

An alternative basis of $Z[\mathbb{C}[\mathfrak{S}_n]]$ is given by idempotents. These are given by the formula
\begin{equation}
    \label{eq:idemps def}
    \mathcal{E}_\lambda := \frac{\dim \lambda}{|\lambda|!}\sum_{\mu \vdash n} \chi^\lambda(\mu) C_\mu.
\end{equation}

Idempotents have four notable properties. 
\begin{enumerate}
    \item \textbf{Orthogonal idempotents.}
    The elements $\mathcal E_\lambda$, $\lambda\vdash n$, form a basis of $Z[\mathbb C[\mathfrak S_n]]$, and for $\lambda,\lambda'\vdash n$
    \begin{equation}
        \label{eq:Idmeps_orth}
        \mathcal{E}_\lambda \mathcal{E}_{\lambda^{\prime}}= \begin{cases}\mathcal{E}_\lambda & \lambda=\lambda^{\prime}, \\ 0 & \lambda \neq \lambda^{\prime} .\end{cases}
    \end{equation}
    We consider all $\mathcal E_\lambda$ at once as elements of the algebra $\bigoplus_{n\ge0}\mathbb C[\mathfrak S_n]$, in which elements of different degrees are multiplied componentwise, so that the product of an element of $\mathbb C[\mathfrak S_n]$ with an element of $\mathbb C[\mathfrak S_m]$, $n\ne m$, is zero. Then \eqref{eq:Idmeps_orth} holds for all partitions $\lambda,\lambda'$, whether or not $|\lambda|=|\lambda'|$. (This is not the multiplication in $\mathbb C[\mathfrak S_{m}]$ for the standard inclusion $\mathfrak S_n\subset\mathfrak S_m$, where such a product does not vanish in general.)

    \item \textbf{Coefficients.}
    Any element $\theta \in Z[\mathbb{C}[\mathfrak{S}_n]]$ can be written as $\theta=\sum_{\lambda\vdash n}\theta_\lambda\mathcal E_\lambda$. By \eqref{eq:Idmeps_orth}, the coefficients are determined by the action of $\theta$ on $\mathcal E_\lambda$:
    \begin{equation}
        \label{eq:Idmeps coefs}
        \theta\,\mathcal{E}_\lambda = \theta_\lambda\,\mathcal{E}_\lambda.
    \end{equation}

    \item \textbf{Eigenvalues via contents.}
    Let $f(x_1,x_2,\dots,x_n)$ be any symmetric polynomial in $n$ variables. Then the element $f(\mathcal{J}_1,\mathcal{J}_2,\dots,\mathcal{J}_n)$ lies in the center $Z[\mathbb{C}[\mathfrak{S}_n]]$. It is shown in \cite{jucys1974symmetric, murphy1981new}, that the eigenvalues are
    \begin{equation}
        \label{eq:YJM elements on idemps}
        f\left(\mathcal{J}_1, \ldots, \mathcal{J}_n\right) \mathcal{E}_\lambda=f\left(\{j-i\}_{(i, j) \in \lambda}\right) \mathcal{E}_\lambda,
    \end{equation}
    where $f\left(\{j-i\}_{(i, j) \in \lambda}\right)$ denotes the value of $f$ when evaluated at the $n$ quantities $j-i$ corresponding to the boxes $(i,j)\in\mathbb{N}^2$ in the Young diagram of $\lambda\vdash n$. These quantities are the \emph{contents} of the associated boxes in the diagram. Figure~\ref{fig:young-contents} displays the contents of the diagram for the partition $\lambda = (4,2,1)$.
    \begin{figure}[htbp]
        \centering
        \begin{tikzpicture}[scale=1.0]
        \draw[step=1, thick] (0,0) grid (4,1);
        \draw[step=1, thick] (0,-1) grid (2,0);
        \draw[step=1, thick] (0,-2) grid (1,-1);
    
        \node at (0.5,0.5) {$0$};
        \node at (1.5,0.5) {$1$};
        \node at (2.5,0.5) {$2$};
        \node at (3.5,0.5) {$3$};
    
        \node at (0.5,-0.5) {$-1$};
        \node at (1.5,-0.5) {$0$};
    
        \node at (0.5,-1.5) {$-2$};
    
        \foreach \x/\j in {0.5/1,1.5/2,2.5/3,3.5/4}
            \node[above=3pt] at (\x,1) {$\j$};
    
        \foreach \y/\i in {0.5/1,-0.5/2,-1.5/3}
            \node[left=5pt] at (0,\y) {$\i$};
    
        \node at (2,1.85) {Column $j$};
        \node[rotate=90] at (-0.9,-0.5) {Row $i$};
    
        \end{tikzpicture}
        \caption{Young diagram of $\lambda=(4,2,1)$ with box contents $c(i,j)=j-i$.}
        \label{fig:young-contents}
    \end{figure}
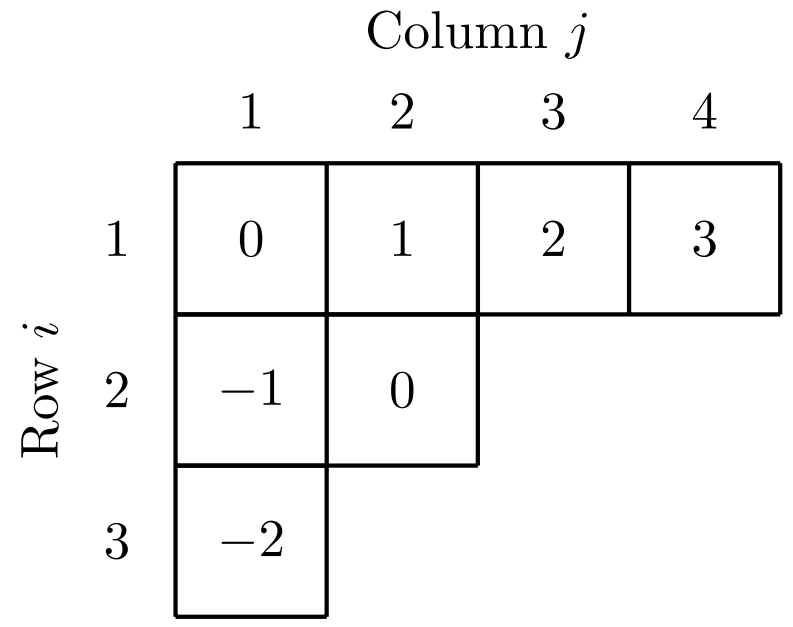

    \item \textbf{Eigenvalues via characters.}
    Another central element is $C_\mu$. As before, this element is diagonal when expressed in the idempotent basis. Its eigenvalues are determined by the definition of $\mathcal{E}_\lambda$ and the first two properties.
    \begin{equation}
        \label{eq:Cmu idemps}
        C_\mu \mathcal{E}_\lambda = \frac{n!}{z_\mu} \frac{\chi^\lambda(\mu)}{\dim \lambda} \mathcal{E}_\lambda := \kappa_\mu(\lambda) \mathcal{E}_\lambda.
    \end{equation}
    This identity yields the inverse change-of-basis matrix, going from the idempotent basis back to the basis of class sums:
    \begin{equation}
        \label{eq:idemps to Cmu}
        C_\mu = \frac{n!}{z_\mu} \sum_{\lambda \vdash n} \frac{\chi^\lambda(\mu)}{\dim \lambda} \mathcal{E}_\lambda.
    \end{equation}

\end{enumerate}

The key object in our proof of Theorem \ref{thm:schur-selberg} is the Frobenius characteristic map $\mathrm{ch}$. It maps the elements of symmetric group algebra $\mathbb{C}[\mathfrak{S}_n]$ to the algebra of symmetric functions of degree $n$, acting linearly \cite[Chapter I, §7]{macdonald1995symmetric}. The rule is as follows.
\begin{equation}
    \label{eq:def Frobenius map}
    \mathrm{ch}[w_{\mu}] = \frac{1}{|\mu|!}p_{\mu}(\mathbf{u}).
\end{equation}

This rule can be extended by linearity to the class sums.
\begin{equation}
    \label{eq:Frobenius map cmu}
    \mathrm{ch}[C_{\mu}] = \frac{1}{z_{\mu}}p_{\mu}(\mathbf{u}).
\end{equation}

We can also use the identity \eqref{eq:idemps def} and linearity. Then
\begin{equation}
    \label{eq:Frobenius map idemps}
    \mathrm{ch}[\mathcal{E}_{\lambda}] = \frac{\dim \lambda }{|\lambda|!} \sum_{\mu \vdash n} \chi^\lambda(\mu) \frac{p_{\mu}(\mathbf{u})}{z_\mu} = \frac{\dim \lambda }{|\lambda|!} s_{\lambda}(\mathbf{u}),
\end{equation}
where we used the connection between power sums $p_\mu$ and Schur functions $s_\lambda$.  
\begin{equation}
\label{eq:p-s-connection}
    \sum_{\mu \vdash n} \chi^\lambda(\mu) \frac{p_{\mu}(\mathbf{u})}{z_\mu} =  s_{\lambda}(\mathbf{u}).
\end{equation}

The partition function $Z_N(\mathbf{u})$ \eqref{eq:partition_function} admits expansions in both the power-sum basis $p_\lambda(\mathbf{u})$ and the Schur basis $s_\lambda(\mathbf{u})$. Consequently, the partition function can be represented via the Frobenius characteristic map. In this language, the relations \eqref{eq:pertfunc power sums} and \eqref{eq:partfunc schur func} can be rewritten as
\begin{equation}
    \label{eq:partfunc Frob map general}
    \begin{aligned}
        Z_N(\mathbf{u}) = \mathrm{ch}\left[ \sum_{\lambda \in \mathcal{P}} \left\langle \prod_{i=1}^{\ell(\lambda)} \tr X^{\lambda_i}\right\rangle C_\lambda \right] = \mathrm{ch}\left[ \sum_{\lambda \in \mathcal{P}} c_{\lambda, N} \frac{|\lambda|!}{\dim \lambda}\mathcal{E}_\lambda \right],
    \end{aligned}
\end{equation}
where the bracketed terms provide two alternative descriptions of the same element in the direct sum of the centers of the symmetric-group algebras, $\bigoplus_{n \ge 0} Z[\mathbb{C}[\mathfrak{S}_n]]$.

For the choice $(p,q)=(0,1)$ (which is the setting studied in \cite{gisonni2021jacobi}), the central element appearing inside the brackets takes the explicit form
\begin{equation}
    \label{eq:central element}
    \sum_{n\ge 0} N^{n} \left(\frac{c_A}{c_A + c_B}\right)^{n} \prod_{i=1}^{n} G\left( \frac{\mathcal{J}_i}{N}\right).
\end{equation}
Consequently, the partition function can be written as
\begin{equation}
    \label{eq:partfunc by Frob map}
    Z_N(\mathbf{u}) = \mathrm{ch} \left[\sum_{n\ge 0} N^{n} \left(\frac{c_A}{c_A + c_B}\right)^{n} \prod_{i=1}^{n} G\left( \frac{\mathcal{J}_i}{N}\right)\right].
\end{equation}

This partition function is a Toda tau function in the times $u_1, u_2,\dots$ \cite{AvM1995, AvM2001, Deift}. In \cite{harnad2015hypergeometric} it was proved that a function $G$ of the form
\begin{equation}
    \label{eq:G_func general}
    G(t) = \frac{\prod_{i=1}^k(1+\gamma_i t)}{\prod_{j=1}^m (1-\delta_j t)}
\end{equation}
yields hypergeometric tau functions via \eqref{eq:partfunc by Frob map}, and these are intimately connected with the theory of the 2D Toda hierarchy. One can choose the parameters $\gamma_i$ and $\delta_j$, so that the general function $G$ \eqref{eq:G_func general} becomes the particular function of interest \eqref{eq:G_func}, and this tau function becomes the partition function of the JUE. 

Our main idea is to express $X_{p,q}$ as an affine transformation of $X_{0,1}$ Any factors and terms generated by this transformation can be shifted to the central element \eqref{eq:central element}. Writing this resulting central element in terms of the class sums $C_\lambda$ makes it possible to compute the correlators and to establish the main theorem \ref{thm:main-theorem}. On the other hand, expressing it in terms of idempotents enables us to determine the coefficients $c_{\lambda, N}$ and to prove theorem \ref{thm:schur-selberg} on the Schur--Selberg type integral.

\section{Proof of Theorem \ref{thm:main-theorem}}
\label{sec:main-thm-proof}

The proof has three steps. First, we use the definition \eqref{eq:Xpq-def} of $X_{p,q}$ through $X_{0,1}$ together with the result of \cite{gisonni2021jacobi}
to express the correlators through coefficients of
$\prod G(\mathcal{J}_i/N)$ in the groups $\mathfrak{S}_D$, $D\le n$
(Section~\ref{subs:reduction}). Second, we lift these coefficients from
$\mathfrak{S}_D$ to $\mathfrak{S}_n$ (Lemmas~\ref{lem:piS} and
\ref{lem:removeS}, Section~\ref{subs:lifting}). Third, we sum over subsets and
collect all dependence on $p,q$ in one central element, which turns out to be
a combination of class sums (Lemma~\ref{lem:classsum},
Section~\ref{subs:central}).

\subsection{Reduction to the case $(p,q)=(0,1)$}
\label{subs:reduction}

The first step is to reduce the correlators of $X_{p,q}$ to those of $X_{0,1}$, which were derived in \cite{gisonni2021jacobi}. For arbitrary real $p,q$ this is built into the definition \eqref{eq:Xpq-def}. The following lemma shows that this definition agrees with the density definition of the ensemble from Section~\ref{sec:intro}.

\begin{lemma}
\label{lem:affine}
Let $\mathrm{Re}\,\alpha,\mathrm{Re}\,\beta>-1$ and let $X_{0,1}\sim\mathbf J(0,1,\alpha,\beta,N)$.
\begin{enumerate}
    \item[(a)] If $p<q$, then $(q-p)X_{0,1}+pI\sim\mathbf J(p,q,\alpha,\beta,N)$.
    \item[(b)] $I-X_{0,1}\sim\mathbf J(0,1,\beta,\alpha,N)$.
    \item[(c)] If $p>q$, then $(q-p)X_{0,1}+pI\sim\mathbf J(q,p,\beta,\alpha,N)$.
\end{enumerate}
\end{lemma}
\begin{proof}
(a) The map $Z\mapsto X=(q-p)Z+pI$ is an affine bijection $\mathcal H_N([0,1])\to\mathcal H_N([p,q])$ with $\dif X=(q-p)^{N^2}\dif Z$, $\det(X-pI)=(q-p)^N\det Z$ and $\det(qI-X)=(q-p)^N\det(I-Z)$. Hence
$\det{}^\alpha Z\,\det{}^\beta(I-Z)\,\dif Z$ is mapped to a constant multiple of $\det{}^\alpha(X-pI)\det{}^\beta(qI-X)\dif X$, and the constant is absorbed by the normalization.

(b) The map $Z\mapsto Z'=I-Z$ is a bijection of $\mathcal H_N([0,1])$ preserving $\dif Z$. Since $\det Z=\det(I-Z')$ and $\det(I-Z)=\det Z'$, the density $\det{}^\alpha Z\,\det{}^\beta(I-Z)$ becomes $\det{}^\beta Z'\,\det{}^\alpha(I-Z')$, which is the density of $\mathbf J(0,1,\beta,\alpha,N)$.

(c) For $p>q$ we have $(q-p)X_{0,1}+pI=qI+(p-q)\,(I-X_{0,1})$. By (b), $I-X_{0,1}\sim\mathbf J(0,1,\beta,\alpha,N)$. Applying (a) to the interval $[q,p]$ with the parameters $(\beta,\alpha)$ gives the claim.
\end{proof}

Throughout the rest of the proof we write $X$ for $X_{p,q}$. By \eqref{eq:Xpq-def}, for any real $p,q$,
\begin{equation}
\left\langle \prod_{j=1}^{\ell(\lambda)} \tr X^{\lambda_{j}} \right\rangle = \left\langle \prod_{j=1}^{\ell(\lambda)} \tr \left((q-p)X_{0,1} + pI\right)^{\lambda_{j}} \right\rangle ,
\end{equation}
and expanding by the binomial theorem we obtain
\begin{equation}        \label{eq:Ti_appear}
    \begin{aligned}
        & \left\langle \prod_{j=1}^{\ell(\lambda)} \tr X^{\lambda_{j}} \right\rangle = \\
        & = \sum_{0 \le T_{i} \le \lambda_{i}} \left(\prod_{i} \binom{\lambda_{i}}{T_{i}} \right) (q-p)^{|T|} p^{|\lambda|-|T|} N^{\ell(\lambda) - \ell(T)} \left\langle \prod_{j=1}^{\ell(T)} \tr X^{T_{j}}_{0,1} \right\rangle,
    \end{aligned}
\end{equation}
where the integers $T_i$ are introduced by the binomial expansion. Since these integers are not necessarily ordered, the tuple $T$ is not, in general, a partition in the strict sense. Nevertheless, we set $|T|=\sum_i T_i$ to be the total sum of its entries, and we let $\ell(T)$ denote the count of indices $i$ for which $T_i\neq 0$. The contribution $N^{\ell(\lambda)-\ell(T)}$ comes from those terms with $T_i=0$, using $\tr X_{0,1}^0=\tr I = N$.

Applying identity \eqref{eq:01cor_G_function}, we obtain
\begin{equation}
    \label{eq:Xpq_G_function_from_01}
    \begin{aligned}
        & \left\langle \prod_{j=1}^{\ell(\lambda)} \tr X^{\lambda_{j}} \right\rangle = \sum_{0 \le T_{i} \le \lambda_{i}} \left(\prod_{i} \binom{\lambda_{i}}{T_{i}} \right) (q-p)^{|T|} p^{|\lambda|-|T|} N^{\ell(\lambda) - \ell(T)} \cdot \\
        & \cdot \sum\limits_{d\ge 0} N^{|T|-d} \left(\frac{c_{A}}{c_{A}+c_{B}}\right)^{|T|} [w_{T} N^{-d}] \prod_{i=1}^{|T|} G\left(\frac{\mathcal{J}_{i}}{N} \right),
    \end{aligned}
\end{equation}
where $w_T$ is a permutation with cycle type $T$, ignoring the order of the $T_i$'s.

\subsection{From $\mathfrak{S}_D$ to $\mathfrak{S}_n$: subsets and the $S$-order}
\label{subs:lifting}

In \eqref{eq:Xpq_G_function_from_01} the coefficient $[w_T]$ is taken in
$\mathfrak{S}_{|T|}$, so different terms live in different groups. To compare
them, and eventually to collect them under one coefficient extraction, we lift
everything to $\mathfrak{S}_n$. For that we sort the sum
$\sum_{0 \le T_i \le \lambda_i}$ by $D=|T|$.

\begin{equation}
    \label{eq:Xpq_D_intro}
    \begin{aligned}
        & \left\langle \prod_{j=1}^{\ell(\lambda)} \tr X^{\lambda_{j}} \right\rangle = \sum_{D=0}^{|\lambda|} (q-p)^{D} p^{|\lambda|-D} \left(\frac{c_{A}}{c_{A}+c_{B}}\right)^{D} \cdot \\
        & \cdot \sum\limits_{d\ge 0} \sum_{0 \le T_{i} \le \lambda_{i}, |T| = D} \left(\prod_{i} \binom{\lambda_{i}}{T_{i}} \right)  N^{D+\ell(\lambda) - \ell(T)-d} [w_{T} N^{-d}] \prod_{i=1}^{D} G\left(\frac{\mathcal{J}_{i}}{N} \right),
    \end{aligned}
\end{equation}

Let $S$ be a subset of $\{1,\dots,|\lambda|\}$ with cardinality $|S| = D$. We denote by $\mathfrak{S}_D^S$ the group of all permutations of the indices in $S$. By placing a bar over the symbol, we obtain $\bar{\mathfrak{S}}_D^S$, which stands for the corresponding subgroup of $\mathfrak{S}_n$.

We define the projector $\pi_S$, which removes all indices $i \notin S$ from a given permutation, ensuring that the resulting permutation lies in $\mathfrak{S}_D^S$. For instance, if $S = \{1,3,4\}$, then $\pi_S (3 5 4)(1 2) = (3 4)(1)$. We further introduce the operator $\bar{\pi}^S$, indicated with a bar, which maps a permutation into $\bar{\mathfrak{S}}_D^S$ by inserting fixed points for the excluded indices. In this case, $\bar{\pi}_S (3 5 4)(1 2) = (3 4)(1)(2)(5)$.

Define the $S$-order on indices as follows. Given a subset $S$, we declare that all indices belonging to $S$ are smaller than all indices not in $S$, while all other pairwise relations between indices remain unchanged. The $i$-th index in this $S$-order will be denoted by $S_i$. We then define $\mathcal{J}_i^S$ to be the YJM element corresponding to the $S$-order:
\[
\mathcal{J}_i^S = \sum_{j<i} (S_j \ S_i).
\]

We can now state the following lemma.
\begin{lemma}
    \label{lem:piS}
    Using the notation introduced above, for any non-negative integer $d$, for every $0 \le D \le n$ and any permutation $w_\lambda$ of cycle type $\lambda$, we have
    \begin{equation}
        \begin{aligned}
            & \sum_{0 \le T_{i} \le \lambda_{i}, |T| = D} \left(\prod_{i} \binom{\lambda_{i}}{T_{i}} \right)  N^{- \ell(T)} [w_{T} N^{-d}] \prod_{i=1}^{D} G\left(\frac{\mathcal{J}_{i}}{N} \right) = \\
            & \sum_{S \subset \{1,...,|\lambda|\}, |S| = D}  N^{- \ell(\pi_S w_\lambda)} [\pi_S w_\lambda N^{-d}] \prod_{i=1}^{D} G\left(\frac{\mathcal{J}_{i}^S}{N} \right).
        \end{aligned}
    \end{equation}
\end{lemma}
\begin{proof}
    To see this, we focus on the following subsets $S$. Fix a permutation $w_\lambda$ and order its cycles from the longest to the shortest. From the cycle of maximal length $\lambda_1$ choose $T_1$ indices, from the cycle of length $\lambda_2$ choose $T_2$ indices, and so on. All the chosen indices are declared to be in $S$, and all remaining indices are excluded from $S$. We denote by $\mathcal{S}(w_\lambda, T)$ the family of subsets $S$ obtained in this way.

    Using this procedure, the subset $S$ can be chosen in $\prod_{i=1}^{\ell(\lambda)} \binom{\lambda_i}{T_i}$ distinct ways. Now consider two different subsets $\tilde S$ and $\hat S$ from $\mathcal{S}(w_\lambda, T)$. It is clear that $\ell(\pi_{\tilde S} w_\lambda) = \ell(\pi_{\hat{S}} w_\lambda)$, since the only distinction between the resulting permutations arises from a relabeling of indices.

    We also can say that
    \begin{equation}
        [\pi_{\hat S} w_\lambda N^{-d}] \prod_{i=1}^{D} G\left(\frac{\mathcal{J}_{i}^{\hat S}}{N} \right) = [\pi_{\tilde S} w_\lambda N^{-d}] \prod_{i=1}^{D} G\left(\frac{\mathcal{J}_{i}^{\tilde S}}{N} \right).
    \end{equation}
    because of the same reason.

    That yields that for given $T$ and $w_\lambda$
    \begin{equation}
        \begin{aligned}
            & \left(\prod_{i} \binom{\lambda_{i}}{T_{i}} \right)  N^{- \ell(T)} [w_{T} N^{-d}] \prod_{i=1}^{D} G\left(\frac{\mathcal{J}_{i}}{N} \right) = \\
            & = \sum_{S \in \mathcal{S}(T,w_\lambda)\}}  N^{- \ell(\pi_S w_\lambda)} [\pi_S w_\lambda N^{-d}] \prod_{i=1}^{D} G\left(\frac{\mathcal{J}_{i}^S}{N} \right).
        \end{aligned}
    \end{equation}
    Summation over $T: |T| = D$ finishes the proof of the lemma, as $\cup_{T:|T|=D} \mathcal{S}(w_\lambda, T) = \cup_{S:|S|=D}S$.
    
\end{proof}

\begin{remark}
    The same argument can be used to prove more general lemma.
    Let $F$ be an arbitrary symmetric function of $D$ variables. Let $\lambda$ be a partition of $n \ge D$. Take a permutation $w_\lambda \in \mathfrak{S}_n$. Take an integer $d$. Then
    \begin{equation}
        \begin{aligned}
            & \sum_{0 \le T_{i} \le \lambda_{i}, |T| = D} \left(\prod_{i} \binom{\lambda_{i}}{T_{i}} \right)  [w_{T} N^{-d}] F\left(\frac{\mathcal{J}_{i}}{N} \right) = \\
            & \sum_{S \subset \{1,...,|\lambda|\}, |S| = D} [\pi_S w_\lambda N^{-d}] F\left(\frac{\mathcal{J}_{i}^S}{N} \right).
        \end{aligned}
    \end{equation}
\end{remark}

Using the lemma \ref{lem:piS}, we obtain
\begin{equation}
    \label{eq:Xpq_piS}
    \begin{aligned}
        & \left\langle \prod_{j=1}^{\ell(\lambda)} \tr X^{\lambda_{j}} \right\rangle = \sum_{D=0}^{|\lambda|} (q-p)^{D} p^{|\lambda|-D} \left(\frac{c_{A}}{c_{A}+c_{B}}\right)^{D} \cdot \\
        & \cdot \sum\limits_{d\ge 0} \sum_{S\subset \{1,\dots,|\lambda|\}, |S| = D}  N^{D+\ell(\lambda) - \ell(\pi_S w_\lambda)-d} [\pi_S w_\lambda N^{-d}] \prod_{i=1}^{D} G\left(\frac{\mathcal{J}_{i}^S}{N} \right).
    \end{aligned}
\end{equation}

The next step is to replace the permutation $\pi_S w_\lambda$ by $w_\lambda$ itself. To this end, we prove the following lemma.

\begin{lemma}
    \label{lem:removeS}
    Let $n=|\lambda|$, for every $0 \le D \le n$ let $S\subset\{1,\dots,n\}$ with $|S|=D$, and let $w_\lambda\in\mathfrak{S}_n$ be a permutation of cycle type $\lambda$. Let $\mathcal{J}_i^S$ be the Young--Jucys--Murphy elements in the $S$-order. Then for every integer $d$
    \begin{equation}
        \label{eq:RemovingS}
        \begin{aligned}
            & \left[\pi_S w_\lambda N^{-d}\right] \prod_{i=1}^{D} G\left(\frac{\mathcal{J}_{i}^S}{N} \right) = \\
            & = \left[w_\lambda N^{-d - n+D + \ell(w_\lambda) - \ell(\pi_S w_\lambda)}\right] \prod_{i=1}^{D} G\left(\frac{\mathcal{J}_{i}^S}{N} \right) \prod_{j=D+1}^n \left(1+\frac{\mathcal{J}_{j}^S}{N}\right).
        \end{aligned}
    \end{equation}
    Equivalently, set
    \begin{equation}
        \label{eq:d_to_tilded}
        \tilde d = d + \ell(\pi_S w_\lambda) - \ell(w_\lambda) - D+n.
    \end{equation}
    Then the left-hand side of \eqref{eq:RemovingS} is the coefficient of $N^{-\tilde d}$ in the coefficient of $w_\lambda$ in the right-hand product. Moreover, only $\tilde d\ge 0$ can contribute.
\end{lemma}
\begin{proof}
    Write $P_S=\prod_{i=1}^{D} G\left(\mathcal{J}_{i}^S/N\right)$ and $Q_S=\prod_{j=D+1}^n \left(1+\mathcal{J}_{j}^S/N\right)$. The statement of the lemma then reads:
    \begin{equation}
        \label{eq:RemovingS_PSQS}
        \left[\pi_S w_\lambda N^{-d}\right] P_S = \left[w_\lambda N^{-d - n+D + \ell(w_\lambda) - \ell(\pi_S w_\lambda)}\right] P_S  Q_S.
    \end{equation}
    The elements $\mathcal{J}_i^S$ with $i\le D$ are sums of transpositions of indices from $S$ only. Hence $P_S$ lies in $\mathbb{C}[\mathfrak{S}_D^S]$, and we can write $P_S=\sum_{\tau\in\mathfrak{S}_D^S} a_\tau\,\tau$, where the coefficients $a_\tau$ are Laurent series in $N^{-1}$.

    We begin by describing how a single bracket in $Q_S$ acts on a chosen permutation. Let $\rho$ be a permutation of the indices $S_1,\dots,S_{m-1}$ with $m>D$, and let $\bar \rho$ be a permutation $\rho$ with $S_m$ being a fixed point, so $\bar \rho$ is a permutation of indices $S_1,\dots,S_{m}$. Then
    \begin{equation}
    \rho\left(1+\frac{\mathcal{J}_m^S}{N}\right)=\bar \rho+\frac1N\sum_{j<m} \bar\rho\,(S_j\ S_m).
    \end{equation}
    The summand $\bar \rho$ corresponds to introducing $S_m$ as an additional 1-cycle. For each term $\bar \rho\,(S_j\ S_m)$, one obtains the permutation $\rho$ with $S_m$ inserted into the cycle that contains $S_j$, placed immediately after $S_j$. Thus, every permutation $\rho'$ of $S_1,\dots,S_m$ appearing on the right-hand side is uniquely determined by requiring that, after deleting $S_m$, it reduces to $\rho$; conversely, each such permutation occurs exactly once in this expansion. Forming a new cycle has weight $1$, whereas inserting $S_m$ into a pre-existing cycle comes with weight $N^{-1}$.

    Apply this insertion  successively for $m=D+1,\dots,n$. For every $\tau\in\mathfrak{S}_D^S$,
    \begin{equation}
        \tau\, Q_S=\sum_{\sigma\in\mathfrak{S}_n:\ \pi_S\sigma=\tau} N^{-(n-D)+(\ell(\sigma)-\ell(\tau))}\,\sigma .
    \end{equation}
    Indeed, there are $n-D$ insertion steps. The number of steps that create a new cycle equals $\ell(\sigma)-\ell(\tau)$ (note that $\sigma$ and $\tau$ belong to symmetric groups of different degrees), because inserting into an already existing cycle does not change the number of cycles. Thus, only the other steps contribute a factor of $N^{-1}$.

We focus on the permutation $\tau=\pi_S w_\lambda$. Multiplying $\pi_S w_\lambda$ by $Q_S$ produces permutations of the form $\pi_S^{-1}[\pi_S w_\lambda]$, each with an associated factor. Among them, one obtains $w_\lambda$ itself. The coefficient of $w_\lambda$ in the expansion of $P_SQ_S$ is
\begin{equation}
    \begin{aligned}
        &[w_\lambda]\,P_SQ_S = [w_\lambda]\,\sum_{\tau}a_\tau \tau Q_S= \\
        & = \sum_{\tau}a_\tau\,[w_\lambda]\,(\tau Q_S)
    = a_{\pi_S w_\lambda}\, N^{-(n-D)+\ell(w_\lambda)-\ell(\pi_S w_\lambda)} = \\
        & = N^{-(n-D)+\ell(w_\lambda)-\ell(\pi_S w_\lambda)} [\pi_S w_\lambda] P_S.
    \end{aligned}
\end{equation}
Taking the coefficient of $N^{-m}$ on both sides, where
$m=d+n-D-\ell(w_\lambda)+\ell(\pi_S w_\lambda)=\tilde d$,
yields
$[w_\lambda N^{-\tilde d}]P_SQ_S=[\pi_S w_\lambda N^{-d}]P_S$,
which is exactly \eqref{eq:RemovingS}.

    Finally, the largest power of $N$ appearing in $P_SQ_S$ is $N^0$, since each factor is of the form $1+O(N^{-1})$. Therefore, $\tilde d\ge 0$.
\end{proof}

Applying Lemma~\ref{lem:removeS} to every term of \eqref{eq:Xpq_piS} and replacing $\tilde d$ by $d$ (which runs over nonnegative integers), we obtain
\begin{equation}
    \label{eq:Xpq_w_lambda factors not in coef}
    \begin{aligned}
        & \left\langle \prod_{j=1}^{\ell(\lambda)} \tr X^{\lambda_{j}} \right\rangle = \sum_{D=0}^{|\lambda|} (q-p)^{D} p^{|\lambda|-D} \left(\frac{c_{A}}{c_{A}+c_{B}}\right)^{D} \sum_{S\subset \{1,\dots,|\lambda|\}, |S| = D}\\
        & \sum\limits_{d\ge 0}   N^{n- d} [w_\lambda N^{-d}] \prod_{i=1}^{D} G\left(\frac{\mathcal{J}_{i}^S}{N} \right)\prod_{j=D+1}^n \left(1+\frac{\mathcal{J}_{j}^S}{N}\right).
    \end{aligned}
\end{equation}

\subsection{Collecting the dependence on $p,q$ in a central element}
\label{subs:central}

Equation \eqref{eq:Xpq_w_lambda factors not in coef} is the outcome of the
previous subsection. For every $D$ and $S$ the coefficient is now extracted at
the same permutation $w_\lambda\in\mathfrak{S}_n$ and at the same power of $N$.
Hence all factors involving $p,q$ can be moved under the coefficient
extraction. This is the moment when we move the factors that appeared because
of the shift from $X_{0,1}$ to $X$ into the central element.
\begin{equation}
    \label{eq:Xpq_w_lambda}
    \begin{aligned}
        & \left\langle \prod_{j=1}^{\ell(\lambda)} \tr X^{\lambda_{j}} \right\rangle = \sum\limits_{d\ge 0}  N^{n- d} [w_\lambda N^{-d}] \sum_{D=0}^{|\lambda|}\sum_{S\subset \{1,\dots,|\lambda|\}, |S| = D} \\
        & (q-p)^{D} p^{|\lambda|-D} \left(\frac{c_{A}}{c_{A}+c_{B}}\right)^{D} \prod_{i=1}^{D} G\left(\frac{\mathcal{J}_{i}^S}{N} \right) \prod_{j=D+1}^n \left(1+\frac{\mathcal{J}_{j}^S}{N}\right).
    \end{aligned}
\end{equation}
Our next and final step is to simplify this expression.

Note that from the definition of $G$ \eqref{eq:G01_initial_expansion_of_permutations} and from the fact that for any $S$ $\prod_{j=1}^n\left(1+a\frac{\mathcal{J}_{j}^S}{N}\right) = \prod_{j=1}^n\left(1+a \frac{\mathcal{J}_{j}}{N}\right)$ one obtains
\begin{equation}
    \begin{aligned}
        & \prod_{i=1}^D G\left(\frac{\mathcal{J}_{i}^S}{N} \right) \prod_{j=D+1}^n \left(1+\frac{\mathcal{J}_{j}^S}{N}\right) = \\
        & = \prod_{i=1}^D \frac{\left(1+\frac{\mathcal{J}_{i}^S}{N} \right)\left(1+\gamma\frac{\mathcal{J}_{i}^S}{N}\right)}{\left(1- \delta\frac{\mathcal{J}_{i}^S} {N} \right)} \prod_{j=D+1}^n\left(1+\frac{\mathcal{J}_{j}^S}{N}\right) = \\
        & = \prod_{i=1}^n \left(1+\frac{\mathcal{J}_i}{N}\right) \prod_{j=1}^D \frac{\left(1+\gamma\frac{\mathcal{J}_j^S}{N}\right) }{\left(1- \delta\frac{\mathcal{J}_j^S}{N}\right)}=\\
        & = \prod_{i=1}^n \frac{\left(1+\frac{\mathcal{J}_i}{N}\right)}{\left(1- \delta \frac{\mathcal{J}_i}{N}\right)} \prod_{j=1}^D \left(1+\gamma\frac{\mathcal{J}_j^S}{N}\right) \prod_{k=D+1}^n \left(1- \delta\frac{\mathcal{J}_k^S}{N}\right)
    \end{aligned}
\end{equation}

Using that, we can rewrite \eqref{eq:Xpq_w_lambda} as
\begin{equation}
    \label{eq:Xpq_less_S}
    \begin{aligned}
        & \left\langle \prod_{j=1}^{\ell(\lambda)} \tr X^{\lambda_{j}} \right\rangle = \sum\limits_{d\ge 0}  N^{n- d} [w_\lambda N^{-d}]\prod_{i=1}^n \frac{\left(1+\frac{\mathcal{J}_i}{N}\right)}{\left(1- \delta \frac{\mathcal{J}_i}{N}\right)} \sum_{D=0}^{|\lambda|}\sum_{S\subset \{1,\dots,|\lambda|\}, |S| = D} \\
        & (q-p)^{D} p^{|\lambda|-D} \left(\frac{c_{A}}{c_{A}+c_{B}}\right)^{D}  \prod_{j=1}^D \left(1+\gamma\frac{\mathcal{J}_j^S}{N}\right) \prod_{k=D+1}^n \left(1- \delta\frac{\mathcal{J}_k^S}{N}\right).
    \end{aligned}
\end{equation}

We introduce the following notation.
\begin{equation}
    \begin{aligned}
        & \Phi := \sum_{D=0}^{n}\sum_{S\subset \{1,\dots,n\}, |S| = D} (q-p)^{D} p^{n-D} \left(\frac{c_{A}}{c_{A}+c_{B}}\right)^{D} \cdot \\
        & \cdot \prod_{j=1}^D \left(1+\gamma\frac{\mathcal{J}_j^S}{N}\right) \prod_{k=D+1}^n \left(1- \delta\frac{\mathcal{J}_k^S}{N}\right).
    \end{aligned}
\end{equation}
The object $\Phi$ is obviously central, as it is symmetric under the exchange of indices. In order to simplify $\Phi$ we establish the following lemma.

\begin{lemma}
    \label{lem:classsum}
    Let $n\ge 0$, $\gamma=1/c_A$ and $\delta=-1/(c_A+c_B)$. With the notation given above,
    \begin{equation}
        \label{eq:C_nu}
        \Phi = \sum_{\nu \vdash n}   N^{\ell(\nu)-n} C_\nu \frac{\prod_{i=1}^{\ell(\nu)} (c_A q^{\nu_i} + c_B p^{\nu_i})}{(c_A+c_B)^{n}}.
    \end{equation}
\end{lemma}
\begin{proof}
    Fix $\nu\vdash n$ and a permutation $w_\nu$ with cycles $w_\nu^1,\dots,w_\nu^{\ell(\nu)}$. We compute the coefficient $[w_\nu]\Phi$ and show that it depends only on $\nu$. $\Phi$ is central, so one can write $\Phi=\sum_{\nu\vdash n}([w_\nu]\Phi)\,C_\nu$.

    Fix $S$ with $|S|=D$. By the argument following \eqref{eq:symmetric-product}, the indices of a cycle $w_\nu^i$ arise successively in the expansion of the product of the factors $(1+a\mathcal{J}_j^S/N)$, with a unique choice of terms for each permutation. Let $D_i=|S\cap w_\nu^i|$. Recall that in the $S$-order all indices of $S$ are lower than all other indices.
    \begin{itemize}
        \item If $D_i\ge1$, the cycle is built by choosing $1$ from the factor of its first $S$-index, then $D_i-1$ transpositions from the factors $(1+\gamma\mathcal{J}^S_j/N)$, then $\nu_i-D_i$ transpositions from the factors $(1-\delta\mathcal{J}^S_k/N)$. Together with the prefactors, this contributes
        \begin{equation}
            f(\nu_i,D_i)=\frac{1}{N^{\nu_i-1}}\gamma^{D_i-1}(-\delta)^{\nu_i-D_i}(q-p)^{D_i}p^{\nu_i-D_i}\left(\frac{c_A}{c_A+c_B}\right)^{D_i}.
        \end{equation}
        \item If $D_i=0$, the cycle is built by choosing $1$ for its first index and $\nu_i-1$ transpositions from the factors $(1-\delta\mathcal{J}_k^S/N)$. This contributes
        \begin{equation}
            g(\nu_i)=\frac{1}{N^{\nu_i-1}}(-\delta)^{\nu_i-1}p^{\nu_i}.
        \end{equation}
    \end{itemize}
    Here $p^{n-D}$ and $(q-p)^D\left(\frac{c_A}{c_A+c_B}\right)^D$ were split among the cycles, since $\sum_i(\nu_i-D_i)=n-D$ and $\sum_i D_i=D$. Therefore the contribution of a single $S$ to $[w_\nu]\Phi$ is
    \begin{equation}
        \prod_{i:\,D_i=0} g(\nu_i)\prod_{i:\,D_i\ge1} f(\nu_i,D_i).
    \end{equation}

    The contribution depends on $S$ only via $(D_1,\dots,D_{\ell(\nu)})$. Once the $D_i$'s are fixed, there are $\prod_i\binom{\nu_i}{D_i}$ subsets $S$ that correspond to these values. Summing over all $S$ (i.e.\ over all $D$ and all $S$ with $|S|=D$) therefore factorizes:
    \begin{equation}
        \begin{aligned}
            & [w_\nu]\Phi= \sum_{D=0}^n \sum_{\substack{0 \le D_i \le \nu_i \\ \sum D_i = D}} \prod_{i:\,D_i=0} g(\nu_i)\prod_{i:\,D_i\ge1} f(\nu_i,D_i)  \prod_j\binom{\nu_j}{D_j}=\\
            & = \prod_{i=1}^{\ell(\nu)}\left(\sum_{D_i=1}^{\nu_i}\binom{\nu_i}{D_i}f(\nu_i,D_i)+g(\nu_i)\right).
        \end{aligned}
    \end{equation}
    The right-hand side depends only on $\nu$.

    Since $\gamma=1/c_A$ and $-\delta=1/(c_A+c_B)$, the binomial theorem gives
    \begin{equation}
        \label{eq:ith_factor_expression}
        \begin{aligned}
            & \left(\sum\limits_{D_i=1}^{\nu_i} \binom{\nu_{i}}{D_{i}} f(\nu_{i}, D_{i})\right) + g(\nu_i) = \\
            & = \frac{1}{N^{\nu_i-1}}(c_A+c_B)^{-\nu_i}(c_A q^{\nu_i} + c_B p^{\nu_i}),
        \end{aligned}
    \end{equation}

    Multiplying over $i$ and using $\sum_i\nu_i=n$,
    \begin{equation}
        [w_\nu]\Phi=N^{\ell(\nu)-n}\,\frac{\prod_{i=1}^{\ell(\nu)}(c_Aq^{\nu_i}+c_Bp^{\nu_i})}{(c_A+c_B)^n},
    \end{equation}
    which gives \eqref{eq:C_nu}.
\end{proof}

Using this lemma, we continue to transform the expression \eqref{eq:Xpq_less_S}.
\begin{equation}
    \label{eq:Xpq_final_inter_classes}
    \begin{aligned}
        & \left\langle \prod_{j=1}^{\ell(\lambda)} \tr X^{\lambda_{j}} \right\rangle = \sum\limits_{d\ge 0}  N^{n- d} [w_\lambda N^{-d}]\prod_{i=1}^n \frac{\left(1+\frac{\mathcal{J}_i}{N}\right)}{\left(1- \delta \frac{\mathcal{J}_i}{N}\right)} \cdot \\
        & \cdot \sum_{\nu \vdash n}   N^{\ell(\nu)-n} C_\nu \frac{\prod_{i=1}^{\ell(\nu)} (c_A q^{\nu_i} + c_B p^{\nu_i})}{(c_A+c_B)^{n}}.
    \end{aligned}
\end{equation}
As in Section~\ref{subs:yjm} (see the discussion after \eqref{eq:G01_initial_expansion_classes}), the factors $\left(1+\frac{\mathcal{J}_i}{N}\right)$ and $\left(1- \delta \frac{\mathcal{J}_i}{N}\right)^{-1}$ can be rewritten as follows. 
\begin{equation}
    \label{eq:factors_to_sums}
    \begin{aligned}
        &\prod_{i=1}^n \frac{\left(1+\frac{\mathcal{J}_i}{N}\right)}{\left(1- \delta \frac{\mathcal{J}_i}{N}\right)} = \\
        & = \left(\sum_{\mu \vdash n}N^{\ell(\mu)-n} C_\mu \right) \left(\sum_{r\ge 0}N^{-r}\delta^r \sum_{1 \le a_1 \le a_2 \le \dots \le a_r \le n}\mathcal{J}_{a_1}\dots \mathcal{J}_{a_r}\right),
    \end{aligned}
\end{equation}
After multiplying these factors into the sum over $\nu$, we obtain
\begin{equation}
    \label{eq:Xpq_final_classes}
    \begin{aligned}
        & \left\langle \prod_{j=1}^{\ell(\lambda)} \tr X^{\lambda_{j}} \right\rangle = \sum\limits_{d\ge 0}  N^{n- d} [w_\lambda N^{-d}] \left(\sum_{\mu \vdash n}N^{\ell(\mu)-n} C_\mu \right)\cdot \\
        & \cdot  \left(\sum_{r\ge 0}N^{-r}(-c_A - c_B)^{-r} \sum_{1 \le a_1 \le a_2 \le \dots \le a_r \le n}\mathcal{J}_{a_1}\dots \mathcal{J}_{a_r}\right) \cdot \\
        & \cdot \sum_{\nu \vdash n}   N^{\ell(\nu)-n} C_\nu \frac{\prod_{i=1}^{\ell(\nu)} (c_A q^{\nu_i} + c_B p^{\nu_i})}{(c_A+c_B)^{n}}.
    \end{aligned}
\end{equation}
Once more, the overall exponent of $\frac1N$ is
\begin{equation}
    d = n-\ell(\mu) + n - \ell(\nu) + r.
\end{equation}
Moreover, the Riemann--Hurwitz relation takes the form
\begin{equation}
    2g-2 = -2n + n -\ell(\mu) + n - \ell(\nu) + n - \ell(\lambda) + r.
\end{equation}
As noted following \eqref{eq:G01_initial_expansion_classes}, the coefficient of $w_\lambda N^{2-2g - n - \ell(\lambda)}$ in the product involving these conjugacy classes and the YJM elements equals $h_g(\lambda, \mu ,\nu)/ |[\lambda]|$.

That allows us to obtain the final expression.
\begin{equation}
    \begin{aligned}
        & \left<\prod_{j=1}^{\ell(\lambda)}\tr X^{\lambda_{j}}\right> = \sum\limits_{g \in \mathbb{Z}} \frac{(-1)^{|\lambda|}}{N^{2g-2+ \ell(\lambda)}}\cdot\\
        & \cdot \sum\limits_{\mu, \nu \vdash |\lambda|} \frac{\prod_{i=1}^{\ell(\nu)}\left( c_A q^{\nu_{i}} + c_B p^{\nu_{i}} \right)}{(-c_{A} - c_{B})^{2g-2 + \ell(\lambda) + \ell(\mu) + \ell(\nu)}} \frac{h_{g}(\lambda, \mu, \nu)}{|[\lambda]|}. 
    \end{aligned}
\end{equation}
This completes the proof of Theorem~\ref{thm:main-theorem}. \qed

The intermediate formulas \eqref{eq:Xpq_w_lambda} and \eqref{eq:Xpq_final_classes} will be used again in the next section.

\section{Corollaries and special cases of Theorem \ref{thm:main-theorem}}
\label{sec:particular-p-q}

In this section we examine several distinguished choices of $(p,q)$. The
results of this section follow from Theorem~\ref{thm:main-theorem}, or from
the intermediate formulas of its proof, and serve as consistency checks as
well as applications: the degenerate case $p=q$ (Section~\ref{subs:p=q}), the
Jacobi ensembles with $\{p,q\}=\{0,1\}$ (Section~\ref{subs:mirror}), the zero-mean ensemble
(Section~\ref{subs:zero-mean}), and its Gaussian limit (Section~\ref{subs:GUE-limit}).

\subsection{The degenerate case $p=q$}
\label{subs:p=q}

First, suppose that $p=q$. By definition \eqref{eq:Xpq-def}, $X_{p,p}=pI$. Hence, for $n=|\lambda|$, one simply has
\begin{equation}
    \label{eq:X=pI}
    \prod_{j=1}^{\ell(\lambda)}\tr X^{\lambda_j}
    =p^nN^{\ell(\lambda)}.
\end{equation}
Thus, all correlators are non-random. This result is recovered directly from the equation \eqref{eq:Xpq_w_lambda}, which is repeated below.
\begin{equation}
    \label{eq:Xpq_w_lambda_repeat}
    \begin{aligned}
        & \left\langle \prod_{j=1}^{\ell(\lambda)} \tr X^{\lambda_{j}} \right\rangle = \sum\limits_{d\ge 0}  N^{n- d} [w_\lambda N^{-d}] \sum_{D=0}^{|\lambda|}\sum_{S\subset \{1,\dots,|\lambda|\}, |S| = D} \\
        & (q-p)^{D} p^{|\lambda|-D} \left(\frac{c_{A}}{c_{A}+c_{B}}\right)^{D} \prod_{i=1}^{D} G\left(\frac{\mathcal{J}_{i}^S}{N} \right) \prod_{j=D+1}^n \left(1+\frac{\mathcal{J}_{j}^S}{N}\right).
    \end{aligned}
\end{equation}
In the case $p=q$ all the terms with non-zero values of $D$ can be omitted. Thus, the correlators are
\begin{equation}
    \label{eq:p=q_inter}
    \begin{aligned}
        & \left\langle \prod_{j=1}^{\ell(\lambda)} \tr X^{\lambda_{j}} \right\rangle = \sum\limits_{d\ge 0}  N^{n- d} [w_\lambda N^{-d}]  p^{|\lambda|}  \prod_{j=1}^n \left(1+\frac{\mathcal{J}_{j}}{N}\right) = \\
        &  = \sum\limits_{d\ge 0}  N^{n- d} p^{|\lambda|}[w_\lambda N^{-d}] \sum_{\nu \vdash n}    N^{\ell(\nu)-n} C_\nu = \\
        & = p^{n}N^{\ell(\lambda)},
    \end{aligned}
\end{equation}
which agrees with the expected result in \eqref{eq:X=pI}.

\subsection{The standard Jacobi ensemble and its mirror image}
\label{subs:mirror}

Next, consider the choice $q=1$, $p=0$. This is the standard Jacobi unitary ensemble considered in \cite{gisonni2021jacobi}. Indeed, in this case $\prod_{i=1}^{\ell(\nu)}\left( c_A q^{\nu_{i}} + c_B p^{\nu_{i}} \right)=c_A^{\ell(\nu)}$, and Theorem~\ref{thm:main-theorem} reproduces \eqref{GGR-correlators}.

Now consider the mirror choice $q=0$, $p=1$. By \eqref{eq:Xpq-def}, $X_{1,0}=I-X_{0,1}$, and by Lemma~\ref{lem:affine}(b) $X_{1,0}\sim\mathbf J(0,1,\beta,\alpha,N)$. Therefore the correlators of $X_{1,0}$ are obtained from those of $X_{0,1}$ by the interchange $\alpha\leftrightarrow\beta$, that is $c_A\leftrightarrow c_B$. Applying this interchange to \eqref{GGR-correlators} (the denominator $(-c_A-c_B)$ is symmetric and only the numerator changes), we obtain
\begin{equation}
    \label{GGR-correlators_repeat_B}
    \begin{aligned}
        & \left<\prod_{j=1}^{\ell(\lambda)}\tr X^{\lambda_{j}}_{1,0}\right> = \\
        & = \sum_{g\in \mathbb{Z}}\frac{(-1)^{|\lambda|}}{N^{2g-2+\ell(\lambda)}}\sum_{\mu,\nu\vdash|\lambda|}\frac{c_{B}^{\ell(\nu)}\, h_{g}(\lambda,\mu,\nu)}{(-c_{A}-c_{B})^{\ell(\mu)+\ell(\nu)+\ell(\lambda)+2g-2}\, |[\lambda]|}.
    \end{aligned}
\end{equation}
On the other hand, Theorem~\ref{thm:main-theorem} with $(p,q)=(1,0)$ gives the same expression: since $q^{\nu_i}=0$ and $p^{\nu_i}=1$, the factor $\prod_{i}(c_Aq^{\nu_i}+c_Bp^{\nu_i})$ equals $c_B^{\ell(\nu)}$, and nothing else in \eqref{eq:theorem_statement} depends on $(p,q)$. The proof of Theorem~\ref{thm:main-theorem} uses only the definition \eqref{eq:Xpq-def} of $X_{p,q}$ in terms of $X_{0,1}$, so it applies without change to $p>q$. Thus the two derivations (the symmetry argument and Theorem~\ref{thm:main-theorem}) agree, which is a consistency check of Theorem~\ref{thm:main-theorem}.

\subsection{The zero-mean ensemble and parity vanishing}
\label{subs:zero-mean}

Next, consider the zero-mean case. The mean of the eigenvalue distribution vanishes when $q=c_B,\ p=-c_A$. For this choice, the factor associated with a part $\nu_i$ becomes
$c_Ac_B^{\nu_i}+c_B(-c_A)^{\nu_i}$. Equivalently, it can be written as
$c_Ac_B(c_B^{\nu_i-1}-(-c_A)^{\nu_i-1})$. Therefore, if $\nu_i=1$, this factor vanishes. Consequently, only partitions $\nu$ with no parts equal to $1$ can contribute to the correlators in the zero-mean case. 

There is an important special symmetric case of this zero-mean ensemble when $c_A=c_B$. Since the zero-mean condition requires $q=c_B$ and $p=-c_A$, in this case we have $q=-p=c_A=c_B$. The factor corresponding to a part $\nu_i$ then becomes
\begin{equation}
    c_A^{\nu_i+1}\left(1+(-1)^{\nu_i}\right).
\end{equation}
Hence, it vanishes whenever $\nu_i$ is odd and is nonzero whenever $\nu_i$ is even. It follows that, for $c_A=c_B$ in the zero-mean case, a partition $\nu$ contributes only if all of its parts are even. The immediate consequence is that the correlators of $X_{-c_A,c_B}$ are zero whenever $|\lambda|$ is odd.

This is the case we use in the study of limit shapes (see Section~\ref{sec:discussion}).

\subsection{The Gaussian limit}
\label{subs:GUE-limit}

The zero-mean ensemble with $c_A=c_B$ admits another important limiting case. Let $c_A=c_B\to\infty$ and set
$q=\xi c_B, p=-\xi c_A, \xi=\sqrt{2/c_A}$. In this limit the correlators of $X=X_{p,q}$ converge, coefficient by coefficient in the expansion in powers of $N^{-1}$, to the correlators of the Gaussian unitary ensemble.

\begin{cor}
    \label{cor:GUE}
    In the regime described above, the correlators of the matrices $X$ exhibit the following limiting genus expansion.

    \begin{equation}
        \label{eq:Xpq_final_GUE_lemma}
        \begin{aligned}
            & \left\langle \prod_{j=1}^{\ell(\lambda)} \tr X^{\lambda_{j}} \right\rangle \sim \\
            & \sim \sum\limits_{g \in \mathbb{Z}} \frac{1}{N^{2g-2+\ell(\lambda)}}  [w_\lambda N^{-\left(2g-2+\ell(\lambda) + n\right)}] \left(\sum_{\mu \vdash n}N^{\ell(\mu)- \frac{3n}{2}} C_\mu \right) C_{(2,\dots,2)}. 
        \end{aligned}
    \end{equation}
\end{cor}

\begin{proof}
Indeed, we begin by recalling equation \eqref{eq:Xpq_final_classes}. 
\begin{equation}
    \label{eq:Xpq_final_classes_repeat}
    \begin{aligned}
        & \left\langle \prod_{j=1}^{\ell(\lambda)} \tr X^{\lambda_{j}} \right\rangle = \sum\limits_{d\ge 0}  N^{n- d} [w_\lambda N^{-d}] \left(\sum_{\mu \vdash n}N^{\ell(\mu)-n} C_\mu \right)\cdot \\
        & \cdot  \left(\sum_{r\ge 0}N^{-r}(-c_A - c_B)^{-r} \sum_{1 \le a_1 \le a_2 \le \dots \le a_r \le n}\mathcal{J}_{a_1}\dots \mathcal{J}_{a_r}\right) \cdot \\
        & \cdot \sum_{\nu \vdash n}   N^{\ell(\nu)-n} C_\nu \frac{\prod_{i=1}^{\ell(\nu)} (c_A q^{\nu_i} + c_B p^{\nu_i})}{(c_A+c_B)^{n}}.
    \end{aligned}
\end{equation}
Let us use the identities $c_A = c_B$, $q = \xi c_B, p = - \xi c_A$. Also we reflect the fact that $\nu$ is a partition with even parts by $\nu \vdash_2 n$ (with $n$ being even).
\begin{equation}
    \label{eq:Xpq_final_classes_repeat_qp}
    \begin{aligned}
        & \left\langle \prod_{j=1}^{\ell(\lambda)} \tr X^{\lambda_{j}} \right\rangle = \sum\limits_{d\ge 0}  N^{n- d} [w_\lambda N^{-d}] \left(\sum_{\mu \vdash n}N^{\ell(\mu)-n} C_\mu \right)\cdot \\
        & \cdot  \left(\sum_{r\ge 0}N^{-r}(-2c_A)^{-r} \sum_{1 \le a_1 \le a_2 \le \dots \le a_r \le n}\mathcal{J}_{a_1}\dots \mathcal{J}_{a_r}\right) \cdot \\
        & \cdot \sum_{\nu \vdash_2 n}   N^{\ell(\nu)-n} C_\nu \frac{\xi^n}{2^n} (2c_A)^{\ell(\nu)}.
    \end{aligned}
\end{equation}
The right-hand side forms a series in $c_A$. The leading asymptotic contribution comes from $r=0$ and from taking $\ell(\nu)$ as large as possible. Recall that all parts of $\nu$ are even, hence the smallest part is $2$. Therefore, the partition with maximal length is $(2,\ldots,2)$, whose length is $n/2$. As a result, the dominant term reads
\begin{equation}
    \label{eq:Xpq_final_classes_repeat_main_entry}
    \begin{aligned}
        & \left\langle \prod_{j=1}^{\ell(\lambda)} \tr X^{\lambda_{j}} \right\rangle \sim \\
        & \sim \sum\limits_{d\ge 0}  \frac{\xi^n}{2^{\frac n2}} (c_A)^{\frac n2} N^{n- d} [w_\lambda N^{-d}] \left(\sum_{\mu \vdash n}N^{\ell(\mu)- \frac{3n}{2}} C_\mu \right) C_{(2,\dots,2)} .
    \end{aligned}
\end{equation}
The Riemann–Hurwitz formula reads
\begin{equation}
    \label{eq:RH_GUE}
    2g-2 = \frac{n}{2} - \ell(\lambda) - \ell(\mu).
\end{equation}
With $\xi$ defined above the prefactor $\xi^n c_A^{n/2}2^{-n/2}$ equals $1$. Using that, we rewrite the summation over $d$ as a summation over $g$. With this change of variables, we recover the standard expression for the correlators of GUE random matrices \eqref{eq:Xpq_final_GUE_lemma}, whose eigenvalues follow the classical Wigner semicircle law.
\begin{equation}
    \label{eq:Xpq_final_GUE}
    \begin{aligned}
        & \left\langle \prod_{j=1}^{\ell(\lambda)} \tr X^{\lambda_{j}} \right\rangle \sim \\
        & \sim \sum\limits_{g \in \mathbb{Z}} \frac{1}{N^{2g-2+\ell(\lambda)}}  [w_\lambda N^{-\left(2g-2+\ell(\lambda)+ n \right)}] \left(\sum_{\mu \vdash n}N^{\ell(\mu)- \frac{3n}{2}} C_\mu \right) C_{(2,\dots,2)} 
    \end{aligned}
\end{equation}
\end{proof}

\begin{remark}
Equation \eqref{eq:Xpq_final_GUE_lemma} is classical (see, e.g., \cite{Borot_2020, okounkov2000random}) and is tightly connected to the interpretation of the Gaussian ensemble in terms of polygon gluings. The permutation $F:=w_\lambda$ encodes $\ell(\lambda)$ polygons (\emph{faces}) having $\lambda_i$ edges, with edges labeled according to $w_\lambda$. A permutation $E$ in the class $C_{(2,2,\dots,2)}$ represents a choice of pairings of \emph{edges} to be glued. Finally, the permutation $V$ coming from the class sum $C_\mu$ records $\ell(\mu)$ \emph{vertices} of the resulting surface: each cycle of $w_\mu$ corresponds to a vertex. The Euler characteristic can be easily computed with the data presented by these permutations:
\begin{equation}
    \chi = \underbrace{\ell(\mu)}_{\#\text{vertices}}-\underbrace{n/2}_{\#\text{pairs of edges}} + \underbrace{\ell(\lambda)}_{\#\text{faces}},
\end{equation}
which is exactly \eqref{eq:RH_GUE}.

The standard condition guaranteeing that the polygon gluing is consistently defined is
\begin{equation}
    F \cdot E \cdot V = \mathrm{Id}. 
\end{equation}
We refer to \cite[Section 2]{Borot_2020} for further details.
\end{remark}

\section{Proof of Theorem \ref{thm:schur-selberg}}
\label{sec:thm-SS-proof}

In this section we prove Theorem~\ref{thm:schur-selberg}. It does not depend
on Section~\ref{sec:particular-p-q}. We use the same central element as in the
proof of Theorem~\ref{thm:main-theorem}, but expand it in the basis of
idempotents $\mathcal{E}_\lambda$ instead of the class sums $C_\nu$.

We may now write the partition function for general $(p,q)$ by applying the Frobenius characteristic map together with formula \eqref{eq:Xpq_final_inter_classes}:
\begin{equation}
    \label{eq:part func result}
    \begin{aligned}
        & Z_N(\mathbf{u}) = \mathrm{ch} \Bigg[\sum_{n\ge0} \frac{N^{n}}{(c_A+c_B)^{n}} \prod_{i=1}^n \frac{\left(1+\frac{\mathcal{J}_i}{N}\right)}{\left(1- \delta \frac{\mathcal{J}_i}{N}\right)} \cdot \\
        & \cdot \sum_{\nu \vdash n}   N^{\ell(\nu)-n} C_\nu \prod_{i=1}^{\ell(\nu)} (c_A q^{\nu_i} + c_B p^{\nu_i})\Bigg].
    \end{aligned}
\end{equation}

Next, we expand the central element (i.e., the argument of $\mathrm{ch}$) in the idempotent basis. Denoting this element by $\theta$, we determine its coefficients $\theta_\lambda$ using the second idempotent property \eqref{eq:Idmeps coefs}:
\begin{equation}
    \begin{aligned}
        & \theta_\lambda \mathcal{E}_\lambda = \Bigg[\sum_{n\ge0} \frac{N^{n}}{(c_A+c_B)^{n}} \prod_{i=1}^n \frac{\left(1+\frac{\mathcal{J}_i}{N}\right)}{\left(1- \delta \frac{\mathcal{J}_i}{N}\right)} \cdot \\
        & \cdot \sum_{\nu \vdash n}   N^{\ell(\nu)-n} C_\nu \prod_{i=1}^{\ell(\nu)} (c_A q^{\nu_i} + c_B p^{\nu_i})\Bigg] \mathcal{E}_\lambda.
    \end{aligned}
\end{equation}
We note that only $n = |\lambda|$ contributes. Let the class sums $C_\nu$ act on $\mathcal{E}_\lambda$. Since this action is diagonal, it produces the eigenvalues $\kappa_\nu(\lambda)$ introduced in \eqref{eq:Cmu idemps}:
\begin{equation}
    \begin{aligned}
        & \theta_\lambda \mathcal{E}_\lambda = \frac{N^{n}}{(c_A+c_B)^{n} } \Bigg[\prod_{i=1}^n \frac{\left(1+\frac{\mathcal{J}_i}{N}\right)}{\left(1- \delta \frac{\mathcal{J}_i}{N}\right)} \Bigg] \cdot \\
        &  \cdot \sum_{\nu \vdash n}   N^{\ell(\nu)-n} \kappa_\nu(\lambda) \prod_{i=1}^{\ell(\nu)} (c_A q^{\nu_i} + c_B p^{\nu_i})\mathcal{E}_\lambda.
    \end{aligned}
\end{equation}
After that, we apply the Young--Jucys--Murphy elements to $\mathcal{E}_\lambda$. This action is likewise diagonal, and its eigenvalues are expressed via the contents, as in \eqref{eq:YJM elements on idemps}:
\begin{equation}
    \begin{aligned}
        & \theta_\lambda \mathcal{E}_\lambda = \frac{N^{n}}{(c_A+c_B)^{n} } \left(\prod_{(i,j) \in \lambda} \frac{1+ N^{-1} (j-i)}{1-\delta N^{-1} (j-i)}\right) \cdot \\
        &  \cdot \sum_{\nu \vdash n}   N^{\ell(\nu)-n} \kappa_\nu(\lambda) \prod_{i=1}^{\ell(\nu)} (c_A q^{\nu_i} + c_B p^{\nu_i})\mathcal{E}_\lambda.
    \end{aligned}
\end{equation}
Finally, substituting the definition of $\kappa_\nu(\lambda)$ from \eqref{eq:Cmu idemps}, using $\delta = - 1/(c_A+c_B)$, and introducing $\varepsilon = N^{-1}$, we arrive at
\begin{equation}
    \begin{aligned}
        & \theta_\lambda = \frac{1}{(c_A+c_B)^{n} } \left(\prod_{(i,j) \in \lambda} \frac{1+ \varepsilon (j-i)}{1+ \frac{\varepsilon}{c_A+c_B} (j-i)}\right) \cdot \\
        &  \cdot \sum_{\nu \vdash n}   \varepsilon^{-\ell(\nu)} \frac{n!}{z_\nu} \frac{\chi^\lambda(\nu)}{\dim \lambda} \prod_{i=1}^{\ell(\nu)} (c_A q^{\nu_i} + c_B p^{\nu_i}).
    \end{aligned}
\end{equation}

We also recall the relation between $\theta_\lambda$ and $c_{\lambda, N}$ given in \eqref{eq:partfunc Frob map general}:
\begin{equation}
    \label{eq:clambda result}
    \begin{aligned}
        & c_{\lambda, N} = \theta_\lambda \frac{\dim \lambda}{|\lambda|!} = \frac{1}{(c_A+c_B)^{n} } \left(\prod_{(i,j) \in \lambda} \frac{1+ \varepsilon (j-i)}{1+ \frac{\varepsilon}{c_A+c_B} (j-i)}\right) \cdot \\
        &  \cdot \sum_{\nu \vdash n}   \varepsilon^{-\ell(\nu)}  \frac{\chi^\lambda(\nu)}{z_\nu}\prod_{i=1}^{\ell(\nu)} (c_A q^{\nu_i} + c_B p^{\nu_i}).
    \end{aligned}
\end{equation}

Let us also recall definition \eqref{eq:Schur-Selberg-like-integral}:
\begin{equation}
    I_{\lambda} = \int_{(p,q)^{N}} \chi_{\lambda}(x) \Delta^{2}(x)\prod_{a=1}^{N}(q-x_{a})^{\beta} (x_{a}-p)^{\alpha} \dif^{N}x.
\end{equation}
By \eqref{eq:clambda by Ilambda}, the coefficient $c_{\lambda, N}$ coincides with the ratio $I_\lambda/I_\varnothing$. Combining this identification with \eqref{eq:clambda result} yields the proof of Theorem \ref{thm:schur-selberg}:
\begin{equation}
    \begin{aligned}
        &\frac{I_\lambda}{I_{\varnothing}}  = \left(\prod_{(i,j) \in \lambda} \frac{1+ \varepsilon (j-i)}{1+ \frac{\varepsilon}{c_{A}+c_{B}} (j-i)}\right) \cdot \\
        & \cdot \sum\limits_{\nu \vdash n} \frac{\chi^{\lambda}(\nu)}{z_{\nu}} \frac{\varepsilon^{-\ell(\nu)}}{(c_{A}+c_{B})^n}\prod_{k=1}^{\ell(\nu)} (q^{\nu_{k}} c_{A}+ p^{\nu_{k}}c_B).
    \end{aligned}
\end{equation}

\section{Conclusion and open problems}
\label{sec:discussion}

We have extended the genus expansion of \cite{gisonni2021jacobi} from the
standard Jacobi ensemble to $\mathbf J(p,q,\alpha,\beta,N)$ with arbitrary
$p,q$ (Theorem~\ref{thm:main-theorem}) and obtained the Schur--Selberg type
integral on $[p,q]$ (Theorem~\ref{thm:schur-selberg}). In both results the
endpoints enter only through the multiplicative factor
$\prod_i (c_A q^{\nu_i}+c_B p^{\nu_i})$. In this section we explain why the
dependence on $(p,q)$ is so simple and formulate several open problems.

\paragraph{Structural interpretation.}
Let $H$ be a rational function of one variable, let $a=(a_k)_{k\ge1}$ be a sequence,
and put $a_\nu=\prod_i a_{\nu_i}$. Consider the central element
\begin{equation}
    \label{eq:concl_central}
    \sum_{n\ge0}\ \prod_{i=1}^{n}H\!\left(\frac{\mathcal J_i}{N}\right)
    \sum_{\nu\vdash n}a_\nu\, C_\nu .
\end{equation}
By \eqref{eq:YJM elements on idemps} and \eqref{eq:Cmu idemps} it acts on
$\mathcal E_\mu$, $\mu\vdash n$, by the scalar
\begin{equation}
    \prod_{\square\in\mu}H\!\left(\frac{c(\square)}{N}\right)\cdot
    \frac{n!}{\dim\mu}\sum_{\nu\vdash n}a_\nu\frac{\chi^\mu(\nu)}{z_\nu},
\end{equation}
where $c(\square)=j-i$ is the content of the box $\square=(i,j)$. One can choose the specialization $\mathbf u^*$, so that $p_k(\mathbf u^*) = a_k$, and use the relation between symmetric functions \eqref{eq:p-s-connection} to show that the scalar is
\begin{equation}
    \prod_{\square\in\mu}H\!\left(\frac{c(\square)}{N}\right)\cdot
    \frac{n!}{\dim\mu}\, s_\mu(\mathbf u^*).
\end{equation}

Applying $\mathrm{ch}$ to \eqref{eq:concl_central} and using \eqref{eq:Frobenius map idemps}, we get
\begin{equation}
    \label{eq:concl_hypergeometric}
    \mathrm{ch}\,[\eqref{eq:concl_central}]
    =\sum_{\mu}r_\mu\, s_\mu(\mathbf u)\, s_\mu(\mathbf u^*),
    \qquad r_\mu=\prod_{\square\in\mu}H\!\left(\frac{c(\square)}{N}\right),
\end{equation}
which is the hypergeometric tau function of the 2D Toda hierarchy in the
sense of Orlov and Shcherbin \cite{orlov2001hypergeometric}, with the second set of times $\mathbf u^*$
specialized to $p_k(\mathbf u^*)=a_k$.

The partition function of the Jacobi ensemble has this form. The weight $G$
of the body of the paper factorizes as $G(t)=(1+\gamma t)\,H(t)$ with
\[
H(t)=\frac{1+t}{1-\delta t},\qquad \gamma=\frac1{c_A},\quad \delta=-\frac1{c_A+c_B}.
\]
The factor $\prod_i(1+\gamma\varepsilon\mathcal J_i)$ is absorbed into
$s_\mu(\mathbf u^*)$. For $(p,q)=(0,1)$ this gives the specialization of
$\mathbf u^*$ to an alphabet of $N+\alpha$ copies of $1/(c_A+c_B)$. The
transition from $[0,1]$ to $[p,q]$ (Lemma~\ref{lem:classsum}) leaves $H$
unchanged and deforms this specialization to
\begin{equation}
    \label{eq:concl_specialization}
    p_k(\mathbf u^*)=N\,\frac{c_Aq^k+c_Bp^k}{(c_A+c_B)^k},
\end{equation}
that is, to the power sums of an alphabet consisting of $N+\alpha$ copies of
$q/(c_A+c_B)$ and $N+\beta$ copies of $p/(c_A+c_B)$. Thus the interval
$[p,q]$ affects only the second set of times, while the weight $r_\mu$ is
the same for all $p,q$. This explains the factorized form of
\eqref{eq:theorem_statement} and \eqref{eq:theorem_schur-selberg-type}.

\paragraph{Open problems.}
\begin{enumerate}
    \item For which coefficients $a_\nu$, not necessarily multiplicative, is
    $\mathrm{ch}\,[\eqref{eq:concl_central}]$ a tau function of the KP
    hierarchy in $\mathbf u$? Multiplicativity of $a_\nu$ is sufficient, but we do
    not know whether it is necessary.
    \item Can this construction produce tau functions outside the 2D Toda
    class? 
    \item Which central elements of the form \eqref{eq:concl_central} arise
    from unitary-invariant matrix ensembles? By Lemma~2.3 of
    \cite{gisonni2021jacobi}, every ensemble with a general weight on an
    interval defines a central element, but we do not know how to
    characterize the resulting class.
\end{enumerate}

\paragraph{Applications.}
The zero-mean case of Section~\ref{subs:zero-mean} will be used in our forthcoming work on
limit shapes of Young diagrams and the largest row, in the spirit of
Okounkov's result for the Gaussian unitary ensemble \cite{okounkov2000random}.

\section*{Acknowledgments}
The author thanks A.~Nazarov for numerous fruitful discussions. This article developed from our joint work on limit shapes of Young diagrams, and the problems we investigated together naturally gave rise to the ideas presented here. The author also thanks P. Nikitin and M. Karev for useful conversations and valuable feedback on earlier versions of this manuscript. This work was supported by the Ministry of Science and Higher Education of the Russian Federation (agreement 075-15-2025-344 dated 29/04/2025 for Saint Petersburg Leonhard Euler International Mathematical Institute at PDMI RAS).

\bibliographystyle{halpha}
\bibliography{literature}

\end{document}